\documentclass[11pt]{article}
\usepackage{amscd}
\usepackage{amsmath}
\usepackage{amsfonts}
\usepackage{theorem}
\usepackage{mathrsfs}
\usepackage{rotating}
\usepackage{pictex}
\usepackage{epic}

\input QED.sty

{\theorembodyfont{\slshape}
\newtheorem{theorem}{Theorem}[section]
\newtheorem{proposition}[theorem]{Proposition}
\newtheorem{lemma}[theorem]{Lemma}
\newtheorem{corollary}[theorem]{Corollary}
}
{\theorembodyfont{\rmfamily}
\newtheorem{definition}[theorem]{Definition}
\newtheorem{problem}{Problem}
\newtheorem{remark}[theorem]{Remark}
\newtheorem{example}{Example}

}

\def\N{\mathbb{N}}
\def\Z{\mathbb{Z}}
\def\R{\mathbb{R}}
\def\Q{\mathbb{Q}}
\def\C{\mathbb{C}} 
\def\F{\mathbb{F}} 
\def\P{\mathbb{P}} 
\def\proj{\mathbb{P}} 
\def \Ri{\R^\infty} 

\renewcommand{\a}{\alpha}
\newcommand{\balpha}{\boldsymbol{\alpha}}

\def\FR{{\mathscr F}_{\R}}
\def\FRbit{{\mathscr F}^0_{\R}}

\def\FC{{\mathscr F}_{\C}}

\def\scC{{\mathscr C}}
\def\mZ{{\mathcal Z}}
\def\mI{{\mathcal I}}
\def\mH{{\mathscr H}}

\newcommand{\x}{\times}

\renewcommand{\>}{\rangle}

\renewcommand{\tilde}{\widetilde}

\def\grad{{\rm grad}\,}
\def\size{{\rm size}}
\def\Oh{{\cal O}}
\def\co{{\rm co}}
\def\poly{\scriptstyle{\sf poly}}
\def\BP{{\rm BP}}
\def\mchi{\chi^*}
\def\PW{{\it PW}}
\newcommand{\bm}{{\scriptscriptstyle {\rm BM}}}

\def\algorithm{\begin{center}
               \begin{minipage}{6in}
               \begin{tabbing}
               \marks}
\def\falgorithm{\end{tabbing}
                \end{minipage}
                \end{center}}
\def\marks{nn\= nn\= nn\= nn\= nn\= nn\= nn\= \kill}

\def\P{{\sf P}}
\def\FP{{\sf FP}}
\def\NP{{\sf NP}}

\def\PSPACE{{\sf PSPACE}}
\def\FPSPACE{{\sf FPSPACE}}

\def\RP{{\sf RP}}

\def\PH{{\sf PH}}

\def\CP{{\#\P}}

\def\GCC{{\sf GCC}}
\def\GCR{{\sf GCR}}

\def\PR{{\rm P}_{\kern-1pt\R}}
\def\Padd{{\rm P}_{\mathrm{add}}}
\def\PC{{\rm P}_{\kern-1pt\C}}
\def\NPR{{\rm NP}_{\kern-1pt\R}}
\def\NPC{{\rm NP}_{\kern-2pt\C}}
\def\DNPR{{\rm DNP}_{\kern-1pt\R}}
\def\DNPC{{\rm DNP}_{\kern-2pt\C}}
\def\PAR{{\rm PAR}_{\kern-1pt\R}}
\def\PHR{{\rm PH}_{\kern-1pt\R}}
\def\DPHR{{\rm DPH}_{\kern-1pt\R}}

\def\FPR{{\rm FP}_{\kern-1pt\R}}
\def\FPC{{\rm FP}_{\kern-1pt\C}}
\def\FPAR{{\rm FPAR}_{\kern-0.4pt\R}}
\def\FPARC{{\rm FPAR}_{\kern-0.4pt\C}}

\def\VNP{{\rm VNP}}
\def\FPARadd{{\rm FPAR}_{\rm add}}

\def\CPRi{{\rm \#P}_{\kern-2pt\R}}
\def\CPCi{{\rm \#P}_{\kern-2pt\C}}

\def\FEASR{{\mbox{\sc Feas}_{\kern-0.5pt\R}}}  
\def\FEASRbit{{\mbox{\sc Feas}^0_{\kern-1pt\R}}}
\def\SAS{{\mbox{\sc SAS}_{\kern-0.5pt\R}}}
\def\SASbit{{\mbox{\sc SAS}_{\kern-1pt\R}^0}}
\def\HNC{{\mbox{\sc HN}_{\kern-1pt\C}}}
\def\HNCbit{{\mbox{\sc HN}^0_{\kern-1pt\C}}}
\def\QASC{{\mbox{\sc QAS}_{\kern-1pt\C}}}

\def\DIMR{{\mbox{\sc Dim}_{\kern-0.5pt\R}}}
\def\DIMC{{\mbox{\sc Dim}_{\kern-0.5pt\C}}}
\def\DIMadd{{\mbox{\sc Dim}_{\kern-0.5pt\add}}}
\def\DIMRbit{{\mbox{\sc Dim}^0_{\kern-0.5pt\R}}}
\def\DIMCbit{{\mbox{\sc Dim}^0_{\kern-0.5pt\C}}}
\def\REACH{{\mbox{\sc Reach}_{\kern-0.5pt\R}}}
\def\REACHbit{{\mbox{\sc Reach}^0_{\kern-0.5pt\R}}}
\def\CREACHbit{{\mbox{\sc CReach}^0_{\kern-0.5pt\R}}}

\def\INDEX{\mbox{\rm INDEX}} 

\def\DEGREE{{\mbox{\sc Degree}}}
\def\DEGREEbit{{\mbox{\sc Degree}^0}}

\def\EULER{{\mbox{\sc Euler}_{\R}}}
\def\EULERbit{{\mbox{\sc Euler}_{\R}^0}} 
\def\MEULER{{{\mbox{\sc Euler}_{\R}^*}}}  
\def\MEULERbit{{\mbox{\sc Euler}_{\R}^{*0}}}
\newcommand{\BETTI}[1]{{{\mbox{\sc Betti}({#1})_{\R}}}}
\newcommand{\MBETTI}[1]{{{\mbox{\sc BM-Betti}({#1})_{\R}}}}
\newcommand{\BETTIbit}[1]{{{\mbox{\sc Betti}({#1})_{\R}^0}}}
\newcommand{\MBETTIbit}[1]{{{\mbox{\sc BM-Betti}({#1})_{\R}^0}}}
\def\CCC{{\mbox{\sc \#CC}_{\R}}} 
\def\CCCbit{{\mbox{\sc \#CC}^0_{\R}}} 

\def\REACH{{\mbox{\sc Reach}_{\kern-0.5pt\R}}}
\def\REACHbit{{\mbox{\sc Reach}^0_{\kern-0.5pt\R}}}
\def\CREACHbit{{\mbox{\sc CReach}^0_{\kern-0.5pt\R}}}
\newcommand{\CBETTIbit}[1]{{\mbox{\sc CBetti}^0_{\R}}({#1})}

\begin{title}
{\Large {\bf Counting Complexity Classes for Numeric Computations II:
Algebraic and Semialgebraic Sets}}
\end{title}
\author{Peter B\"urgisser\thanks{Partially supported by DFG  
grant BU~1371. }\\
Dept.\ of Mathematics\\
Paderborn University\\                    
D-33095 Paderborn\\                      
Germany\\            
e-mail: {\tt pbuerg@upb.de}\\
\and
Felipe Cucker\thanks{Partially supported by City University 
SRG grant 7001558.}\\
Dept.\ of Mathematics\\
City University of Hong Kong\\
83 Tat Chee Avenue, Kowloon\\
Hong Kong\\
e-mail: {\tt macucker@math.cityu.edu.hk}
}

\date{}

\begin{document}
\maketitle

\begin{quote}{\small
{\bf Abstract.\quad} We define counting classes~$\CPRi$ and $\CPCi$ in
the Blum-Shub-Smale setting of computations over the real or complex
numbers, respectively.  The problems of counting the number of
solutions of systems of polynomial inequalities over $\R$, or of
systems of polynomial equalities over $\C$, respectively, 
turn out to be natural complete problems in these classes.  
We investigate to what extent the new counting classes capture 
the complexity of computing basic topological invariants of 
semialgebraic sets (over $\R$) and algebraic sets (over $\C$).  
We prove that the problem of computing the (modified) Euler 
characteristic of semialgebraic sets is $\FPR^{\CPRi}$-complete, and
that the problem of computing the geometric degree of complex algebraic
sets is $\FPC^{\CPCi}$-complete.  We also define new counting
complexity classes in the classical Turing model via taking Boolean
parts of the classes above, and show that the problems to compute the
Euler characteristic and the geometric degree of (semi)algebraic sets
given by integer polynomials are complete in these classes. 
We complement the results in the Turing model by proving, 
for all $k\in\N$, the $\FPSPACE$-hardness of the problem of computing 
the $k$th Betti number of the zet of real zeros of a given integer 
polynomial. This holds with respect to the singular homology as well 
as for the Borel-Moore homology.
}\end{quote}

{\small

}

\section{Introduction} 

The theory of computation introduced by Blum, Shub, and 
Smale in~\cite{blss:89} allows for computations over an 
arbitrary ring $R$. Emphasis was put, however, on the 
cases $R=\R$ or $R=\C$. For these two cases, a major 
complexity result in~\cite{blss:89} exhibited natural 
NP-complete problems, namely, the feasibility of 
semialgebraic or algebraic sets, respectively. Thus, 
the complexity of a basic problem in semialgebraic or 
algebraic geometry was precisely characterized in terms 
of completeness in complexity classes. 

In contrast with discrete\footnote{All along this paper we 
use the words {\em discrete}, {\em classical} or {\em Boolean} to 
emphasize that we are refering to the theory of complexity over 
a finite alphabet as exposed in, e.g., \cite{badg:88,papa:94}.} 
complexity theory, these first 
completeness results were not followed by an avalanche of 
similar results. One may say that, if NP-completeness 
exhibits a single problem with different dresses, the wardrobe 
of that problem in the real or complex settings seems to 
be definitely smaller than that in the discrete setting. 

Also in contrast with discrete complexity theory, very 
little emphasis was put on functional problems. These 
attracted attention at the level of analysis of particular 
algorithms, but structural properties of classes of 
such problems have been hardly studied.
So far, the most systematic approach to study the 
complexity of certain functional problems within 
a framework of computations over the reals
is Valiant's theory of 
$\VNP$-completeness~\cite{buer:00-3,vali:79-3,vali:82}.  
However, the relationship of this theory to the 
more general BSS-setting is, as of today, poorly 
understood. 

A recent departure from the situation above is the 
work focusing on complexity classes related with counting 
problems, i.e., functional problems, whose associated 
functions count the number of solutions of some 
decisional problem. 

In classical complexity theory, counting classes were 
introduced by Valiant in his seminal 
papers~\cite{vali:79-2,vali:79-1}. Valiant defined 
$\CP$ as the class of functions which count the 
number of accepting paths of nondeterministic polynomial time 
Turing machines   
and proved that the computation of the permanent is 
$\CP$-complete. This exhibited an unexpected difficulty 
for the computation of a function whose definition is 
only slightly different to that of the determinant, 
a well-known ``easy'' problem. This difficulty was 
highlighted by a result of Toda~\cite{toda:91} 
proving that 
$\PH\subseteq\P^{\CP}$, i.e., that $\CP$ has at 
least the power of the polynomial hierarchy. 

In the continuous setting, i.e., over the reals,  
counting classes were first defined by Meer 
in~\cite{meer:00}. Here a real version $\CPRi$ of the class 
$\CP$ was introduced, but the existence of complete problems 
for it was not studied\footnote{To distinguish 
between classical and, say, real complexity complexity classes, 
we use the subscript $\R$ to indicate the latter. Also,
to further emphasize this distinction, 
we write the former in {\sf sans serif}.}. Instead, the focus of 
Meer's paper are some logical properties of this class 
(in terms of metafinite model theory). After that, 
in~\cite{bucu:02}, an in-depth study of the properties 
of counting classes over $(\R,+,-,\leq)$ was carried out. 
In this setting, real computations are restricted to those which 
do not perform multiplications and divisions. Main 
results in~\cite{bucu:02} include both structural 
relationships between complexity classes and completeness 
results.

The goal of this paper is to further study $\CPRi$ (and 
its version over the complex numbers, $\CPCi$) following 
the lines of~\cite{bucu:02}. A driving motivation is to 
capture the complexity (in terms of completeness results) 
to compute basic quantities 
of algebraic geometry or algebraic topology 
in terms of complexity classes and completeness results. 
Examples for such quantities are: dimension, cardinality of 
0-dimensional sets, geometric degree, 
multiplicities, number of connected or irreducible components, 
Betti numbers, rank of (sheaf) cohomology groups, Euler 
characteristic, etc. To our best knowledge, 
besides~\cite{bucu:02}, the only known non-trivial 
complexity lower bounds for some of these quantities 
are in~\cite{bach:99,reif:79}.   
For other attempts to characterize the intrinsic complexity 
of problems of algebraic geometry, especially elimination, 
we refer to~\cite{hemo:93,mayr:97,mame:82}. 

Capturing the complexity of some of the above problems will 
help to reduce the contrasts 
we mentioned at the beginning of this introduction. 
\medskip

\noindent
{\bf 1. Counting classes}\qquad 
The class $\#\P$ is defined to be the class of functions 
$f\colon\{0,1\}^\infty\to\N$ for which there exists a
polynomial time Turing machine~$M$ and a polynomial $p$ 
with the property that for all $n\in\N$ and all $x\in\{0,1\}^n$, 
$f(x)$ counts the number of strings $y\in\{0,1\}^{p(n)}$ 
such that $M$ accepts $(x,y)$. 

Replacing Turing machines by BSS-machines over $\R$ in the 
definition above, we get a class of functions
$f:\R^\infty\to\N\cup \{\infty\}$, which we denote by 
$\CPRi$. Thus $f(x)$ counts the number of vectors $y\in\R^{p(n)}$ 
such that $M$ accepts $(x,y)$. Note that this number may be 
infinite, that is, $f(x)=\infty$. In a similar way, one 
defines $\CPCi$. 

Feasibility of Boolean combinations of polynomial equalities 
and inequalities 
and of polynomial equations were proved to be $\NPR$-complete 
problems in~\cite{blss:89}. These problems are denoted by 
$\SAS$ and $\FEASR$ respectively. As one may expect, their 
counting versions $\#\SAS$ and $\#\FEASR$, consisting of counting 
the number of solutions of systems as described above, turn out to be 
complete in $\CPRi$. Similarly, the problem $\#\HNC$ consisting of 
counting the number of complex solutions of systems of polynomial 
equations is complete in $\CPCi$. While we prove these results in 
Section~\ref{counting_R}, one of the goals of this paper is 
to show that other problems, of a basic geometric nature, are 
also complete in these counting classes. 
\medskip

\noindent
{\bf 2. Degree, Euler characteristic and Betti numbers}\qquad 
The study of the zero sets of systems of polynomial equations 
is the subject of algebraic geometry. Classically, these zero 
sets, called algebraic varieties, are considered in $k^n$ for some 
algebraically closed field $k$. A central choice for $k$ is $k=\C$. 
Given an algebraic variety $Z$, a number of quantities are attached 
to it, which describe several geometric features of $Z$. 
Examples of such quantities are dimension and degree. 
Roughly speaking, the degree measures how twisted $Z$ is 
embedded in affine space by, 
more precisely, counting how many intersection points it has 
with generic affine subspaces of a certain well-chosen dimension. 
Not surprisingly, an algebraic variety has degree one if and only 
if it is an affine subspace of $\C^n$. The degree of an algebraic 
variety occurs in many results in algebraic geometry. Maybe the 
most celebrated of them is B\'ezout's Theorem. It also occurs in 
the algorithmics of algebraic geometry~\cite{figm:90,hein:83} 
and in lower bounds results~\cite{bucs:96,stra:73-2}.
\smallskip

The birth of algebraic topology is entangled with more than one 
century of attempts to prove a statement of Euler asserting 
that in a polyhedron, the number of vertices plus the number of 
faces minus the number of edges equals 2 (see~\cite{laka:76} 
for a vivid account of this history). A precise definition of 
a generalization of this sum is today (justly) known with 
the name of Euler characteristic or (justly as well) of 
Euler-Poincar\'e characteristic. 

The Euler characteristic of $X$, denoted by $\chi(X)$, 
is one of the most basic 
invariants in algebraic topology. Remarkably, it 
naturally occurs in many applications in other 
branches of geometry. For instance, in differential 
geometry, where it is proved that a compact, connected, 
differentiable manifold $X$ has a non-vanishing vector 
field if and only if $\chi(X)=0$~\cite[p.~201]{stee:65}. 
Also, in algebraic geometry, a generalization  
of the Euler characteristic (w.r.t.\ sheaf cohomology) 
plays a key role in the Riemann-Roch Theorem for 
non-singular projective varieties~\cite{hirz:65}.  
The Euler characteristic has also played a role in complexity 
lower bounds results. For this purpose, Yao~\cite{ayao:92} introduced 
a minor variation of the Euler characteristic.
This {\em modified Euler characteristic}, denoted $\mchi$, 
has a desirable additivity property and coincides with the usual 
Euler characteristic in many cases, e.g., for compact semialgebraic 
sets and complex algebraic varieties.
\smallskip

The Euler characteristic is invariant under homotopy equivalence and
the modified Euler characteristic is invariant under homeomorphism.
Thus, these quantities are used to prove that certain topological
spaces are not homotopy equivalent or homeomorphic. Yet, there exist
simple examples of pairs of non-equivalent spaces which have the same
Euler characteristic. For instance the spheres $S^1$ and $S^3$ of
dimensions 1 and 3, respectively, satisfy $\chi(S^1)=\chi(S^3)=0$ and
they are not homotopy equivalent.  A more powerful object to
distinguish non-equivalent spaces is the sequence of {\em Betti
numbers}. This is a sequence of non-negative integers $b_k(X)$, 
$k\geq 0$, 
associated to a topological space $X$, invariant under homotopy
equivalence, and satisfying that, if the dimension of $X$ is $d$, then
$b_k(X)=0$ for all $k>d$. The quantity $b_0(X)$ has a very simple
meaning: it is the number of connected components of $X$. Roughly
speaking, for $k\geq 1$, $b_k(X)$ counts the number of $k$-dimensional
holes of $X$. We have $b_0(S^1)=b_0(S^3)=1$, $b_1(S^1)=1$,
$b_1(S^3)=b_2(S^3)=0$, and $b_3(S^3)=1$. This shows that $S^1$ and
$S^3$ are not homotopically equivalent (as one could well expect).
The Euler characteristic and the sequence of Betti numbers are not 
unrelated. One has $\chi(X)=\sum_{k\in\N}(-1)^kb_k(X)$. 

Just as with the Euler characteristic, a version of the Betti numbers 
satisfying an additivity property was introduced by 
Borel and Moore~\cite{bomo:60} for 
locally closed spaces~$X$. These Borel-Moore Betti 
numbers $b_k^\bm(X)$ are invariant under homeomorphisms and 
are related to the modified Euler characteristic as follows:
for locally closed spaces $X$ one has 
$\mchi(X)=\sum_{k\in\N}(-1)^kb_k^\bm(X)$. 
\medskip

\noindent
{\bf 3. Completeness results}\qquad 
A semialgebraic subset of $\R^n$ is defined by a Boolean 
combination of 
polynomial equalities and inequalities. Machines over $\R$ decide 
(in bounded time) sets 
which, when restricted to a fixed dimension $n$, are 
semialgebraic subsets of $\R^n$. Therefore, this kind of sets 
are also the natural input of geometric problems in this setting.  
We have already remarked that deciding emptyness of a semialgebraic 
set is $\NPR$-complete, and that counting the number of points 
of such a set is $\CPRi$-complete. One of the main results in 
this paper is that the problem $\MEULER$ consisting of computing 
the modified Euler characteristic 
of a semialgebraic set is $\FPR^{\CPRi}$-complete. 
The class $\FPR^{\CPRi}$ is an extension of $\CPRi$ in which 
we allow a polynomial time 
computation with an oracle (i.e., a black box) for a function~$f$  
in $\CPRi$. This enhances the power of $\CPRi$ by allowing one 
to compute several values of $f$ instead of only one. 
\smallskip

Over the complex numbers, the situation is similar. Natural inputs 
for geometric problems are quasialgebraic sets, i.e., sets defined 
by a Boolean combination of polynomial equations. Of particular 
interest are  
algebraic varieties. We already remarked that deciding emptyness of 
an algebraic variety is $\NPC$-complete and that counting the number 
of points of such a set is $\CPCi$-complete. Another of the main 
results in this paper is that the problem $\DEGREE$ consisting 
of computing the degree of an algebraic variety is 
$\FPC^{\CPCi}$-complete. 
\smallskip

The proofs of our completeness results rely on diverse tools 
drawn from algebraic geometry, algebraic topology, and 
complexity theory. Two of the techniques we use deserve, 
we believe, some highlight. The first one is the use of generic 
quantifiers, describing properties which hold for almost all 
values. A blend of reasonings in logic and geometry  
allows one to eliminate generic quantifiers 
in parameterized formulae. The 
basic idea behind this method appeared already 
in~\cite{hesc:82} and was used also in~\cite{bcss:96}, but 
the method itself was developed 
in~\cite{koir:97-1,koir:99,koir:99a} 
to prove that the problem of computing the dimension of a 
semialgebraic (or complex algebraic) set is complete in $\NPR$ 
(resp. $\NPC$). We extend this method and use this 
in the completeness proofs of both the degree and the Euler 
characteristic problems. 

The second technique we want to highlight is the application 
of Morse theory for the computation of the Euler 
characteristic. The use of Morse functions as an algorithmic 
tool in algebraic geometry goes back to~\cite{grig:88,grvo:88} where 
the ``critical points method'' was developed to decide quantified 
formulae. Several algorithms to compute the Euler characteristic 
of a semialgebraic set reduce first to the case of a smooth 
hypersurface and then apply the fundamental theorem of Morse 
theory~\cite{basu:99,bruc:90,szaf:86}. We proceed similarly. 
It should be noted, however, that our reduction to 
the smooth hypersurface case is different from those in the 
references above since the latter can not be carried out 
within the allowed resources (polynomial time for real 
machines). 
\pagebreak

\noindent
{\bf 4. Completeness results in the Turing model}\qquad 
In the discussion above we considered real solutions of systems of
real polynomials and complex solutions of systems of complex
polynomials. This coincidence between the base field for the space of
solutions and that for the ring of polynomials used to describe 
solution sets
is not necessary. While one may think of several combinations breaking
it, the one that stands out is the consideration of real (or complex)
solutions of polynomial systems over the integers. In practice, the
difference between considering real or integer coefficients in the
input data is reflected in the difference between the numerical
analysis of polynomial systems and their symbolic computation
(computer algebra). Note that if one restricts the input polynomials
for a problem to have integer coefficients, then the input data for
this problem can be encoded in a finite alphabet and may be considered
in the classical setting. To distinguish this discretized version
from its continuous counterpart we will add a superscript ``0'' in the
problem's name. Thus, for instance, $\HNC^0$ is the problem of
deciding the existence of complex solutions of a system of integer
polynomial equations and $\#\HNC^0$ is the problem of counting the
number of these solutions.

The complexity of computer algebra algorithms for, say, $\HNC^0$ 
is described using discrete models of computation (e.g., Turing 
machines). For instance, relatively recent results~\cite{figm:90}  
show that $\HNC^0\in\PSPACE$, and an even more recent result of 
Koiran~\cite{koir:96} shows that, assuming the generalized 
Riemann hypothesis, $\HNC^0\in\RP^{\NP}$. On the other hand, it 
is well-known (and rather trivial) that $\HNC^0$ is $\NP$-hard. 
The complexity of problems like $\FEASRbit$ or $\SASbit$ is much less 
understood, the gap between their known lower $\NP$ and upper 
$\PSPACE$ bounds being much larger. 
 
In this paper we introduce two new counting complexity classes 
in the discrete setting namely, $\GCC$ and $\GCR$. These classes 
are closed under parsimonious reductions and located between 
$\CP$ and $\FPSPACE$. The problem $\#\HNC^0$ is complete in 
$\GCC$ and the problems $\#\SAS^0$ and $\#\FEASR^0$ are complete 
in $\GCR$. In addition, we also prove that $\DEGREEbit$ and 
$\MEULERbit$ are complete in $\FP^{\GCC}$ and 
$\FP^{\GCR}$, respectively, and that 
$\EULERbit$, the problem of computing the (non-modified) 
Euler characteristic of a basic semialgebraic set,  
is complete in $\FP^{\GCR}$. 

Canny~\cite{cann:88} showed that the problem $\CCCbit$ of counting 
the number of connected components of a semialgebraic set described 
by integer polynomials is in $\FPSPACE$. 
On the other hand, a result by Reif~\cite{reif:79,reif:87}  
stating the $\PSPACE$-hardness of a generalized movers problem 
in robotics easily implies the $\FPSPACE$-hardness of the 
problem $\CCCbit$. 

We give an alternative proof of the $\FPSPACE$-hardness of 
$\CCCbit$ following the lines of~\cite{bucu:02}. 
Extending this, we prove that the problem $\BETTIbit{k}$ 
of computing the $k$th Betti number of the real zero set of a 
given integer polynomial is $\FPSPACE$-hard, for fixed $k\in\N$.
We also prove that the problem $\MBETTIbit{k}$ of computing 
the $k$th Borel-Moore Betti number of the set of real zeros of a 
given integer polynomial is $\FPSPACE$-hard. Note that, for 
$k\geq 1$, the membership of $\BETTIbit{k}$ and $\MBETTIbit{k}$ 
to $\FPSPACE$ is, as of today, an open problem.

State-of-the-art algorithmics for computing the Euler 
characteristic or the number of connected components 
of a semialgebraic set suggests that the former is simpler 
than the latter~\cite{basu:99,bapr:99}. 
In a recently published book~\cite[page~547]{bapr:03} it is 
explicitly observed that the Euler characteristic of real algebraic 
sets (which is the alternating sum of the Betti numbers) can be 
currently more efficiently computed than any of the individual 
Betti numbers.

Our results give some explanation for the observed 
higher complexity required for the computation of the number of 
connected components (or higher Betti numbers) compared to the  
computation of the Euler characteristic.  
Namely, $\EULERbit$ is $\FP^{\GCR}$-complete, while 
$\BETTIbit{k}$ is $\FPSPACE$-hard. 
Thus the problem $\BETTIbit{k}$ is not polynomial time 
equivalent to $\EULERbit$ unless there is 
the collapse of complexity classes
$\FP^{\GCR}=\FPSPACE$. 

A similar observation for the Euler characteristic and the 
Betti numbers in the context of semi-linear sets 
and additive machines was made in~\cite[Corollary~5.23]{bucu:02}. 
\medskip

\noindent
{\bf 5. Organization of the paper}\qquad 
We start in Section~\ref{sec:prelim} by recalling basic facts about 
machines and complexity classes over $\R$ and $\C$ as well as about 
semialgebraic and algebraic sets. Then we define in 
Section~\ref{counting_R} the counting complexity classes $\CPRi$ 
and $\CPCi$, introduce different notions of reduction, and prove 
some basic completeness results. 
The technique of generic quantifiers is described in 
Section~\ref{se:gen-quan} and then used in 
Section~\ref{sec:degree} to prove the completeness 
result for $\DEGREE$. The proof of this result is 
preceded by the exposition of some basic facts about smoothness 
and transversality, which lead to a concise way of 
expressing the degree by a parameterized first order formula. 
We prove the completeness of $\MEULER$ in 
Section~\ref{sec:EulerReal} after recalling some basic facts 
from algebraic and differential topology in Section~\ref{se:ADT}.
Section~\ref{se:Turing} deals with complexity in the discrete 
setting. We define the classes $\GCC$ and $\GCR$ and, 
besides some basic completeness results, we prove 
the completeness of $\DEGREEbit$ in $\GCC$ and of 
$\EULERbit$ and $\MEULERbit$ in $\GCR$. 
Finally, we prove the $\FPSPACE$-hardness of the problems
$\BETTIbit{k}$ and $\MBETTIbit{k}$. 
We close the paper in Section~\ref{se:summary} 
with a summary of problems and results, and with some selected 
open problems in Section~\ref{se:open}.
\bigskip

\noindent
{\bf Acknowledgment.\quad} We are thankful to Saugata Basu and 
Pascal Koiran for helpful discussions while writing this paper. 
We are specially indebted to Pascal Koiran since many of 
his results have been a great source of inspiration for us.

\section{Preliminaries about real machines}
\label{sec:prelim}

\subsection{Machines and complexity classes}

We denote by $\Ri$ the disjoint union 
$\Ri = {\bigsqcup}_{n \geq 0}\R^n$, 
where  for $n\ge 0$, $\R^n$ is the standard
$n$-dimensional space over $\R$. 
The space $\Ri$ is a natural one to represent problem 
instances of arbitrarily high dimension. 
For $x\in\R^n\subset\Ri$, we call $n$ the
{\em size} of $x$ and we denote it by $\size(x)$.  
Contained in $\R^\infty$ is the set of bitstrings 
$\{0,1\}^\infty$ defined as 
the union of the sets $\{0,1\}^n$, for $n\in\N$.   

In this paper we will consider BSS-machines 
over $\R$ as they are defined 
in~\cite{bcss:95,blss:89}. Roughly speaking, such a 
machine takes an input from $\Ri$, performs a 
number of arithmetic operations and comparisons 
following a finite list of instructions, and 
halts returning an element in $\Ri$ (or loops forever). 

For a given machine $M$, the function 
$\varphi_M$ associating its output 
to a given input $x\in\Ri$ is called the 
{\em input-output function}.  
We shall say that a function $f:\Ri\to\R^k$, 
$k\leq\infty$, is 
{\em computable} when there is a machine $M$ such that 
$f=\varphi_M$.
Also, a set $A\subseteq\Ri$ is {\em decided} by a machine 
$M$ if its characteristic function $\chi_A:\Ri\to\{0,1\}$ 
coincides with $\varphi_M$. So, for decision problems we 
consider machines whose output space is $\{0,1\}\subset\R$.

We next introduce some central complexity classes.

\begin{definition}\label{def2d1}
A machine $M$ over $\R$ is said {\em to work in polynomial time} when
there is a constant $c\in \N$ such that for every 
input $x\in \Ri$, $M$ reaches its output node after at 
most $\size(x)^c$ steps. The class $\PR$ is then defined as 
the set of all subsets of $\Ri$ that can be accepted by a machine 
working in polynomial time, and the class $\FPR$ as the set 
of functions which can be computed in polynomial time.  
\end{definition}

\begin{definition}
A set $A$ belongs to $\NPR$ if there is a machine $M$ 
satisfying the following condition: for all $x\in\Ri$,  
$x\in A$ iff there exists $y\in\Ri$ such that $M$ 
accepts the input $(x,y)$ within time polynomial 
in $\size(x)$. In this case, the element~$y$ is 
called a {\em witness} for $x$.   
\end{definition}

\begin{remark}
\begin{description}
\item[(i)]
In this model, the element $y$ can be seen as 
the sequence of guesses used in the Turing machine 
model. However, we note that in this definition no 
nondeterministic machine is introduced as a
computational model, and nondeterminism appears 
here as a new acceptance definition for the 
deterministic machine. Also, we note that the length 
of $y$ can be easily bounded by the time bound
$p(\size(x))$. 

\item[(ii)]
Machines over $\C$ are defined as those over $\R$. 
Note, though, that branchings over $\C$ are 
done on tests of the form $z_0=0$. The classes 
$\PC$, $\NPC$, etc., are then naturally defined. 
\end{description}
\end{remark}

In~\cite[Chapter~18]{bcss:95} models for parallel computation  
over $\R$ are defined. Using these models, one defines   
$\PAR$ to be the class of subsets of $\Ri$, whose characteristic 
function can be computed in parallel polynomial time. Also, 
one defines $\FPAR$ to be the class of functions computable in 
parallel polynomial time such that 
$\size(f(x))$ is bounded by a polynomial in $\size(x)$. 
\medskip

\subsection{Algebraic and semialgebraic sets}
\label{se:salg}

Algebraic geometry is the study of zero sets of polynomials 
(or of objects which locally resemble these sets). 
Standard textbooks on algebraic geometry 
are~\cite{hart:77,mumf:76,shaf:74}. 
For information about real algebraic geometry 
we refer to~\cite{beri:90,bocr:87}. 

We very briefly recall some definitions and facts from 
algebraic geometry, which will be needed later on. 

An {\em algebraic set} (or {\em affine algebraic variety})~$Z$
is defined as the zero set 
$$
 Z=\mZ(f_1,\ldots,f_r):=\{x\in\C^n\mid f_1(x)=0,\ldots,f_r(x)=0\}
$$
of finitely many polynomials $f_1,\ldots,f_r\in \C[X_1,\ldots,X_n]$.
The {\em vanishing  ideal} $\mI(Z)$ of $Z$  consists of all the 
polynomials vanishing on $Z$. Note that $\mI(Z)$ might be strictly 
larger than the ideal $I$ generated by $f_1,\ldots,f_r$.
Actually, by Hilbert's Nullstellensatz, $\mZ(I)$ can be 
characterized as the so-called radical of the ideal~$I$.  

A usual compactification of the space $\C^n$ consists 
of embedding $\C^n$ into $\proj^n(\C)$, the {\em projective space} 
of dimension $n$ over $\C$. Recall, this is the set of complex 
lines through the origin in $\C^{n+1}$ and 
$\C^n\hookrightarrow\proj^n(\C)$ maps a point $x\in\C^n$ 
to the line in $\C^{n+1}$ passing through the origin and 
through $(1,x)$. The notion of an affine 
algebraic variety extends to that of a {\em projective variety} by 
replacing polynomials by homogeneous polynomials in 
$\C[X_0,X_1,\ldots,X_n]$, for which elements of $\proj^n(\C)$ 
are natural zeros. The embedding $\C^n\hookrightarrow\proj^n(\C)$ 
extends to the algebraic subsets of $\C^n$ by defining, 
for any such set $Z$, its {\em projective closure} $\overline{Z}$ as 
the smallest projective variety in $\proj^n(\C)$ containing $Z$. 

A {\em basic semialgebraic set} $S\subseteq\R^n$  is defined to be 
a set of the form
$$
 S =\{x\in\R^n\mid g(x)=0, f_{1}(x)>0,\ldots,f_{r}(x)>0\}, 
$$
where $g,f_1,\ldots,f_r$ are polynomials in $\R[X_1,\ldots,X_n]$. 
We say that $S\subseteq \R^n$ is a {\em semialgebraic set} when 
it is a Boolean combination of basic semialgebraic sets in $\R^n$.   
Every semialgebraic set $S$ can be represented as a finite union 
$S=S_1\cup\ldots\cup S_t$ of basic semialgebraic sets.\footnote{This 
respresentation is said to be in {\em Disjunctive  Normal Form}. A 
representation in {\em Conjuntive Normal Form} is defined in the obvious 
manner.}

We will consider algebraic or semialgebraic sets as input data 
for machines over $\R$ or $\C$. These sets are encoded 
by a family of polynomials describing the set as above. 
To fix ideas we will assume, unless otherwise specified,  
that semialgebraic sets are given as unions of basic 
semialgebraic sets. 
So, properly speaking, the input data is not 
the set itself but a description of it. 
Also, we have to define how polynomials themselves are encoded 
as vectors of real (or complex) numbers. 
However, it will turn out that our results have little dependence on 
the choice of the representation of the semialgebraic set 
and on the encoding of the polynomials, cf.~Remark~\ref{re:3repr}. 

A polynomial 
$f = \sum_{e\in I} u_{e}\, x_1^{e_1}\cdots x_n^{e_n}$
is represented in the {\em sparse encoding} by a list of the pairs
$(u_{e},e)$ for $e\in I$, where 
$I=\{e\in\N^n\mid u_e\neq 0\}$.  
The coefficients $u_{e}$ are given as real (or complex) numbers, 
while the exponent vector $e$ is thought to be given by a bit 
vector of length at most $\Oh(n\log\deg f)$. Let $|I|$ be the total 
number of terms and $\delta:=\max\{2,\deg f\}$. Then 
$\size(f):=|I|n \log \delta$ is defined to be the sparse size of $f$. 
The sparse size of a set of polynomials $f_1,\ldots,f_r$ is defined as 
$\sum_{i=1}^r\size(f_i)$.  
To fix ideas, we will always assume that polynomials are given by the  
sparse encoding. 
If we are dealing with integer polynomials~$f$, we will also 
consider their sparse bit size, which is defined 
as the sparse size of~$f$ multiplied by the maximum bit size of 
the occuring integer coefficients. 

We remark that another way of encoding polynomials is the 
{\em dense encoding}. 
Here, a polynomial of degree $d$ in $n$ variables is given 
by the list of its ${n+d\choose d}$ coefficients, which has 
therefore the size of this combinatorial number. Yet another 
way is to encode the polynomial by a {\em straight-line program} 
computing it, cf.\ \cite{bcss:95,bucs:96}. In this case, the size 
of the encoding of $f$ is the length of the straight-line 
program. 

\subsection{Some known completeness results}

We first recall the basic notions of reduction for 
classes of decision problems.\footnote{This definition is 
actually for a class $\cal C$ containing $\NPR\cap\co\NPR$. To define 
$\PR$-completeness, a stronger notion of reduction is necessary.}  

\begin{definition}
\begin{enumerate}
\item Let $S,T\subseteq\Ri$. We say that 
$\varphi\colon\Ri\to\Ri$ is a {\em reduction} from 
$S$ to $T$ if $\varphi$ can be computed in polynomial 
time and, for all $x\in\Ri$, $x\in S$ if and only if $\varphi(x)\in T$. 

\item We say that $S$ {\em Turing reduces to $T$} if there exists 
an oracle machine which, with oracle $T$, decides  
$S$ in polynomial time.  

\item Let ${\cal C}$ be any class of subsets of $\Ri$. 
We say that a set $T$ is {\em hard} for $\cal C$ 
if, for every $S\in{\cal C}$, there is a  
reduction from $S$ to $T$. We say that $T$ is {\em 
$\cal C$-complete} if, in addition, 
$T\in{\cal C}$.

\item The notions of {\em Turing-hardness} or {\em Turing-completeness} 
are defined similarly. 
\end{enumerate}
\end{definition}
The extension of this definition to $\C$ is immediate. 

The following problems describing variants of the basic feasibility 
problem over $\R$ and $\C$ were introduced and studied 
in~\cite{blss:89}.
\begin{description}
\item{$\HNC$} ({\em Hilbert's Nullstellensatz})\quad 
Given a finite set of complex multivariate polynomials, 
decide whether these polynomials have a common complex zero. 

\item{$\FEASR$} ({\em Polynomial feasibility})\quad 
Given a real multivariate polynomial, decide whether it has a real root. 

\item{$\SAS$} ({\em Semialgebraic satisfiability})\quad 
Given a semialgebraic set $S$, decide whether it is nonempty.
\end{description}

In~\cite{blss:89}, the following fundamental 
completeness result was proved.

\begin{theorem}\label{th:bss}
The problem $\HNC$ is $\NPC$-complete and the 
problems $\FEASR$ and $\SAS$ are $\NPR$-complete.  
\hfill$\Box$
\end{theorem}

Consider the following decision problems related to the 
computation of the dimension of algebraic or semialgebraic sets. 

\begin{description}
\item{$\DIMC$} ({\em Algebraic dimension})\quad 
Given a finite set of complex polynomials
with affine zero set $Z$ and $d\in\N$, decide whether $\dim Z\geq d$.

\item{$\DIMR$} ({\em Semialgebraic dimension})\quad 
Given a semialgebraic set $S$ and $d\in\N$, decide 
whether $\dim S\geq d$.
\end{description}

We denote by $\DIMCbit$ the restriction of the problem~$\DIMC$
to input polynomials with integer coefficients. 
This problem can be encoded in a finite alphabet and may thus be 
studied in the classical Turing setting. 
The problems $\DIMRbit$ and $\HNCbit$ are defined similarly.

Koiran~\cite{koir:97-1,koir:99a} significantly extended 
the list of known geometric $\NPC$- or $\NPR$-complete problems
by showing the following. 

\begin{theorem}\label{th:dim}
\begin{description}
\item[(i)] 
$\DIMC$ is $\NPC$-complete, and $\DIMCbit$ is
equivalent to $\HNC^0$ with respect to 
polynomial-time many-one reductions.
\item[(ii)] 
$\DIMR$ is $\NPR$-complete, and $\DIMRbit$ is 
equivalent to $\FEASR^0$ with respect to 
polynomial-time many-one reductions.
\hfill$\Box$
\end{description}
\end{theorem}

\section{Counting Complexity Classes}
\label{counting_R}

\begin{definition}\label{def1}
We say that a function $f\colon\Ri\to\N\cup\{\infty\}$ 
belongs to the class $\CPRi$ if there exists a 
polynomial time machine~$M$ over~$\R$ and 
a polynomial $p$ such that, for all $x\in\R^n$, 
$$
  f(x)=|\{y\in\R^{p(n)}\mid M\mbox{ accepts } (x,y)\}|.
$$
The complexity class $\FPR^{\CPRi}$ consists of all 
functions $f\colon\Ri\to\Ri$, which can be computed in 
polynomial time using oracle calls to functions in $\CPRi$. 
\end{definition}

\begin{remark}
\begin{description}
\item[(i)]
The class $\CPRi$ is the one defined by 
Meer in~\cite{meer:00}. 
\item[(ii)]
The counting classes $\CPCi$ and $\FPC^{\CPCi}$
are defined mutatis mutandis. Also, replacing $\R$ 
by $\Z_2$ in Definition~\ref{def1} one obtains the 
classical $\CP$. 
\end{description}
\end{remark}

We next define appropriate notions of reduction 
and completeness.

\begin{definition}
\begin{enumerate}
\item Let $f,g\colon\Ri\to\N\cup\{\infty\}$. We say that 
$\varphi\colon\Ri\to\Ri$ is a {\em parsimonious reduction} from 
$f$ to $g$ if $\varphi$ can be computed in polynomial 
time and, for all $x\in\Ri$, $f(x)=g(\varphi(x))$. 

\item We say that $f$ {\em Turing reduces to $g$} if there exists 
an oracle machine which, with oracle $g$, computes 
$f$ in polynomial time.  

\item Let $\cal C$ be $\CPRi$ or $\FPR^{\CPRi}$. 
We say that a function $g$ is {\em hard} for $\cal C$ 
if, for every $f\in{\cal C}$, there is a parsimonious 
reduction from $f$ to $g$. We say that $g$ is {\em 
$\cal C$-complete} if, in addition, $g\in{\cal C}$. 

\item The notions of {\em Turing-hardness} or {\em Turing-completeness} 
are defined similarly. 
\end{enumerate}
\noindent
The extension of this definition to $\C$ is immediate. 
\end{definition}

We define now the following counting versions of the basic 
feasibility problems $\HNC,\FEASR$, and $\SAS$. 

\begin{description}
\item{$\#\HNC$} ({\em Algebraic point counting})\quad 
Given a finite set of complex multivariate polynomials,
count the number of complex common zeros, 
returning $\infty$ if this number is not finite. 

\item{$\#\FEASR$} ({\em Real algebraic point counting})\quad 
Given a real multivariate polynomial,
count the number of its real roots, 
returning $\infty$ if this number is not finite. 

\item{$\#\SAS$} ({\em Semialgebraic point counting})\quad 
Given a semialgebraic set~$S$, compute its 
cardinality if $S$ is finite, and return $\infty$ otherwise.
\end{description}

As was to be expected, these counting problems turn out to be complete 
in the classes $\CPCi$ and $\CPRi$, respectively. In the sequel, 
given $n\in\N$, we denote by $[n]$ the set $\{1,\ldots,n\}$.

\begin{theorem}\label{completeness}
\begin{description}
\item[(i)]
The problem $\#\HNC$ is $\CPCi$-complete 
(with respect to parsimonious reductions).
\item[(ii)]
The problems $\#\FEASR$ and  $\#\SAS$ are $\CPRi$-complete 
with respect to Turing reductions.  
\end{description} 
\end{theorem}

\begin{proof}
For part (i) simply check that the reductions given in the 
corresponding $\NPC$-completeness result by Blum, Shub and 
Smale~\cite{blss:89} (see also~\cite{bcss:95}) are parsimonious. 

The proof of part (ii) requires a more careful look at the 
reduction in~\cite{bcss:95}. In this proof, a machine $M$ solving a 
given problem in $\NPR$ is considered and a reduction is 
established, which associates to every input $\omega\in\Ri$, 
a conjunctive normal form~$\psi_{\omega}$ 
$$
    \bigwedge_{i\in I}\bigg(g_i(x)=0\;\vee\;\bigvee_{j\in J_i} 
       f_{ij}(x)>0\bigg) 
$$ 
(the fact that there is only one equality in each clause is 
achieved by adding squares). An important point to remark here 
is that, while the cardinality of $I$ is bounded by a polynomial 
in the size of $\omega$, the cardinalities $r_i$ of the sets $J_i$ are 
independent of~$\omega$ and depend only on~$M$. 

Now consider one of the clauses of $\psi_{\omega}$ 
\begin{equation}\label{eq:clause}
    g_i(x)=0\;\vee\;\bigvee_{j\in J_i} f_{ij}(x)>0 .
\end{equation}
Considering that  
$g_i$ may be at a point $x$ either $=0$ or $\neq0$, and that 
$f_{ij}$ may be either $<0$, $=0$ or $>0$ we have 
$2\times 3^{r_i}$ possibilities for the signs of 
$g_i,f_{i1},\ldots,f_{ir_i}$ at a point $x$. From them, 
only $K_i=2\times 3^{r_i}-2^{r_i}$ satisfy the clause 
(\ref{eq:clause}). We conclude that we can rewrite this 
clause as an {\em exclusive} disjunction of $K_i$ conjunctions 
of the form  
\begin{equation}\label{eq:clause2}
    g_i(x)\triangle_i 0\;\wedge\;
    \bigwedge_{j\in J_i} f_{ij}(x)\square_{ij}0
\end{equation}
where $\triangle_i\in\{=,\neq\}$ and 
$\square_{ij}\in\{<,=,>\}$. Now replace in (\ref{eq:clause2}) 
the occurrences 
\begin{eqnarray*}
   g_i(x)\neq 0 \qquad&\mbox{by}&\qquad g_i(x)z_i-1=0,\\
   f_{ij}(x)>0 \qquad&\mbox{by}&\qquad f_{ij}(x)y_{ij}^2-1=0,\\
   f_{ij}(x)<0 \qquad&\mbox{by}&\qquad f_{ij}(x)y_{ij}^2+1=0,\\
   f_{ij}(x)=0 \qquad&\mbox{by}&\qquad f_{ij}(x)=0 \wedge y_{ij}^2-1=0.
\end{eqnarray*}
This yields a system of equalities which has, for every solution~$x$ 
of (\ref{eq:clause2}), exactly $2^{r_i}$ solutions in the variables~$x,y,z$. 
Now, for $\ell\in [K_i]$, reduce the system 
in (\ref{eq:clause2}) corresponding to $\ell$ to a 
single equation $F_{i\ell}(x,y,z)=0$ by adding squares and the 
clause (\ref{eq:clause}) to an equation $F^*_i(x,y,z)=0$ by taking 
$F^*_i=\prod_{\ell=1}^{K_i}F_{i\ell}$. Note that, for each solution 
$x$ of $\psi_{\omega}$ there are exactly $2^r$ different solutions 
$(x,y,z)$ of the polynomial
\begin{equation*}
   F:= F^*_1(x,y,z)^2+\cdots+F^*_{m}(x,y,z)^2
\end{equation*} 
where $m$ is the cardinality of $I$ and $r=r_1+\ldots+r_m$. 

The $\CPRi$-Turing-hardness of $\FEASR$ now follows. Finish the reduction 
above by quering $\FEASR$ for the polynomial~$F$ and divide the 
result by $2^k$.  
\end{proof}

\begin{remark}
The proof of Theorem~\ref{completeness} shows that 
the version of $\SAS$ with semialgebraic sets given in 
conjunctive normal form is $\CPRi$-complete with respect to 
parsimonious reductions. 
\end{remark}

\begin{proposition}\label{prop_bounds}
If $f\in\CPRi$ then, for all 
$x\in\R^n$ for which $f(x)$ is finite, 
the bit-size of $f(x)$ is 
bounded by a polynomial in the size of $x$.
\end{proposition}

\begin{proof}
To prove the statement note that, given 
$x\in\Ri$, there exist polynomials $p,q$ such that 
the set of witnesses for $x$ is a semialgebraic subset 
of $\R^{p(n)}$ defined by a union of at most 
$2^{q(n)}$ basic semialgebraic sets, each of them 
described by a system of at most $q(n)$ inequalities 
of polynomials in $p(n)$ variables with degree at 
most $2^{q(n)}$. If this set is finite, its cardinality 
coincides with the number of its connected components.
Now use the bounds for the number of 
connected components of such basic semialgebraic sets 
(see e.g.~\cite[Thm.~11.1]{bucs:96} 
or~\cite[Prop.~7, Chapt.~16]{bcss:95}),  
which follow from the well-known Ole\u\i nik-Petrovski-Milnor-Thom 
bounds~\cite{miln:64,olei:51,olpe:49,thom:65}.
\end{proof} 

We next locate the newly defined counting complexity classes 
within the landscape of known complexity classes. 

\begin{theorem}\label{th:inclusions}
We have $\FPR^{\CPRi} \subseteq\FPAR$.
(To interpret this, represent~$\infty$ by an
element of $\R-\N$.)  
\end{theorem}

\begin{proof}
By Theorem~\ref{completeness}(i), it is sufficient to prove that 
$\#\SAS$ belongs to $\FPAR$. By Theorem~\ref{th:dim}(ii), 
the problem of computing the dimension of a semialgebraic set
is in $\FPR^{\NPR}$, and therefore, in $\FPAR$. 
We use this to compute $\#\SAS$ as follows. 
Given a semialgebraic set, we check whether it is 
zero dimensional. If yes, we 
return its number of connected components, otherwise we 
return $\infty$. This is in $\FPAR$ due to the main 
result in~\cite{bapr:99,grvo:92,hers:94}.
\end{proof}

\begin{remark}
Versions of Proposition~\ref{prop_bounds} and of 
Theorem~\ref{th:inclusions} 
hold over $\C$ as well, with  proofs similar to those over $\R$. 
\end{remark}

The following lemma will be useful later on. 
It is an immediate consequence of the definition of 
the counting classes. 

\begin{lemma}\label{le:addition}
Let $f\colon\Ri\x\{0,1\}^\infty\to\N$ be a function 
in $\CPRi$. Assign to~$f$ and a polynomial $p$ the following 
function $g\colon\Ri\to\N$ obtained by summation: 
for $x\in\R^n$, 
$$
 g(x) =\sum_{y\in\{0,1\}^{p(n)}} f(x,y).
$$
Then $g$ belongs to $\CPRi$. 
A similar statement holds over $\C$.
\hfill$\Box$ 
\end{lemma}

\section{Generic quantifiers}
\label{se:gen-quan}

Our completeness results for $\DEGREE$ and $\MEULER$ 
crucially depend on Koiran's method~\cite{koir:97-1,koir:99,koir:99a} 
to eliminate generic quantifiers in parameterized formulas. 
In this section, we further develop Koiran's method in 
order to adapt it to our purposes. 
The main difference to~\cite{koir:97-1,koir:99,koir:99a} is the 
introduction of the notion of a partial witness sequence 
(compared to the notion of a witness sequence from~\cite{koir:97-1}).  

\subsection{Efficient quantifier elimination over the reals}
\label{se:qe}

For convenience of the reader, we recall a well-known result about 
efficient quantifier elimination over the reals from 
Renegar~\cite[Part III]{rene:92abc}. In the sequel 
$\FR$ denotes the set of first order formulas over the language 
of the theory of ordered fields with constant symbols
for real numbers. The subset of formulas with constant symbols   
for 0 and 1 only, is denoted by $\FRbit$.  

\begin{theorem}\label{pro:qe}
Let $F$ be a formula in $\FRbit$ in prenex form with $k$ free 
variables, $n$~bounded variables, $w$ alternating quantifier blocks, 
and $m$ atomic predicates given by polynomials of degree at most 
$\delta\ge 2$ with integer coefficients of bit-size at most~$\ell$. 
That is, $F$ has the form
$$
 (Q_1 x^{(1)}\in\R^{n_1}) \ldots  (Q_w x^{(w)}\in\R^{n_w})
  G(y,x^{(1)},\ldots,x^{(w)})
$$
with alternating quantifiers $Q_i\in\{\exists,\forall\}$ and 
free variables $y=(y_1,\ldots,y_k)\in\R^k$;
the quantifier free formula $G$ is a Boolean function of $m$ atomic predicates 
$$
 g_j(y,x^{(1)},\ldots,x^{(w)})\Delta_j 0,\quad 1\le j \le m ,
$$
where the $g_j$ are integer polynomials of degree at most $\delta$ and with 
coefficients of bit-size at most $\ell$. Hereby, 
$\Delta_j$ is any of the standard relations
$\{\ge,>,=,\neq,\leq,<\}$. 

Then $F$ is equivalent to a quantifier-free formula $F'$ 
in disjunctive normal form 
$$
  \bigvee_{i=1}^I \bigwedge_{j=1}^{J_i} (h_{ij}\Delta_{ij}0) ,
$$
where $h_{ij}$ are integer polynomials with degree at most $D$ and  
bit-size at most $L$, and such that 
$$
 \log D \le 2^{\Oh(w)}\Big(\prod_{i=1}^w n_i\Big) \log (m\delta),\quad  
 \log L \le 2^{\Oh(w)}\Big(\prod_{i=1}^w n_i\Big) 
 \log (m\delta) + \log (k+\ell). 
$$ 
Moreover, the number $M:=\sum_{i=1}^I J_i$ of atomic predicates 
satisfies the bound 
$$
 \log M \le 2^{\Oh(w)}k\Big(\prod_{i=1}^w n_i\Big) \log (m\delta).\qed
$$ 
\end{theorem}

\subsection{Construction of generic points} 
\label{se:constr-gen-pt}

\begin{definition}
Let $F\in\FR$ have free variables $a_1,\ldots,a_k$. 
We say that $F$ is {\em Zariski-generically true}  
if the set of values $a\in\R^k$ not satisfying $F(a)$ 
has dimension strictly less than~$k$. 
We express this fact by writing $\forall^\ast a\, F(a)$
using the {\em generic universal quantifier} $\forall^\ast$.
\end{definition}

\begin{remark}\label{re:guq}
\begin{description}
\item[(i)]
Let $F\in\FR$ have $k$ free variables 
and coefficient field $K$, i.e., $K$ is the field generated 
by the coefficients of all the polynomials occuring in $F$. 
Then $\forall^\ast a\, F(a)$ is equivalent to each of the 
following statements:
\begin{description}
\item[(a)] $\{a\in\R^k\mid F(a)\}$ is dense in $\R^k$
with respect to the Euclidean topology, 

\item[(b)] 
$\forall \epsilon \in\R\ \forall a\in\R^k\  \exists a'\in \R^k\ 
 \big( \epsilon >0 \Rightarrow F(a') \wedge \|a-a'\| < 
   \epsilon \big)$,

\item[(c)] 
$\forall a\in\R^k \big( \mbox{ $a_1,\ldots,a_k$ 
algebraically independent over $K$ } 
\Longrightarrow F(a) \big)$. 
\end{description}
Part~(b) shows that $\forall^\ast a\, F(a)$ can be expressed by 
a first order formula. Hence by using the generic quantifier 
we still describe semialgebraic sets. 
\item[(ii)]
One can define the {\em generic existential quantifier} $\exists^\ast$ 
by 
$$
   \exists^* a F(a)\equiv \neg\forall^*a\neg F(a).
$$ 
Note that $\exists^* a F(a)$ iff the set of values 
$a\in\R^k$ satisfying $F(a)$ has dimension~$k$. We may say 
that $F$ is {\em Euclidean-generically true}.  
\item[(iii)]
For first order formulas over the language~$\FC$  
of the theory of fields with constant symbols for complex numbers,  
one can define $\forall^*$ and $\exists^*$ just as above. It is not 
difficult to see, however, that  these two quantifiers 
coincide over $\C$. That is, Zariski genericity equals Euclidean genericity.
\end{description}
\end{remark}

The following result was proved in Koiran~\cite[Cor.~1]{koir:99a}. 

\begin{proposition}\label{pro:con-gen-pt}
\begin{description}
\item[(i)]
Let $F\in\FRbit$ be in prenex form with $k$ free variables, 
$n$ bounded variables, $w$ alternating quantifier blocks, and 
$m$ atomic predicates given by polynomials of degree at most 
$\delta\ge 2$ with integer coefficients of bit-size at most~$\ell$. 
If $F$ is Zariski-generically true, then a point $\a\in\Z^k$ 
satisfying~$F$ can be computed by a division-free arithmetic 
straight-line program $\Gamma$ 
of length $\Oh( k n^w \log (m\delta) + \log\ell)$ having
1 as its only constant and no inputs. 
\item[(ii)]
There exists a Turing machine which, with input 
$(k,n,w,m,\delta,\ell)$, computes $\Gamma$ in time polynomial 
in the length of $\Gamma$. This machine does not depend 
on~$F$. 
\end{description}
\end{proposition}

Since we will need the proof method behind this result later on, 
we recall the proof. A first ingredient is the following easy lemma, 
whose proof can be found for instance in~\cite{koir:97}.  

\begin{lemma}\label{le:constr-ts}
For positive integers $k,L,D$ recursively define 
$$
 \alpha_1 := 2^L,\ 
 \alpha_j := 1 + \alpha_1 (D+1)^{j-1}\alpha_{j-1}^D 
 \mbox{ for $2\le j \le k$}. 
$$ 
Then $h(\alpha_1,\ldots,\alpha_k)\ne 0$ for any 
integer polynomial $h$ in $k$ variables of degree at most~$D$ 
and coefficients of absolute value less than $2^L$. 
\hfill$\Box$
\end{lemma}

The sequence $\alpha_1,\ldots,\alpha_k$ in Lemma~\ref{le:constr-ts} 
can be computed by a straight-line program $\Gamma$ performing 
$\Oh(k\log D + \log L)$ arithmetic operations and which has 
$1$ as its only constant. 

\proofof{Proposition~\ref{pro:con-gen-pt}}
Put $S=\{a\in\R^k\mid F(a)\mbox{ holds}\}$.
We use Theorem~\ref{pro:qe} to replace the formula~$F$ by 
an equivalent quantifier free formula 
$
   F'=\bigvee_{i=1}^{I}\bigwedge_{j=1}^{J_i} h_{ij}\Delta_{ij}0 
$
and claim that 
$$
  \bigcap_{i,j}\{a\in\R^k \mid h_{ij}(a) \neq 0\} \subseteq S.
$$ 
Otherwise, there would be some $a\in\R^k- S$ such that 
$h_{ij}(a)\ne 0$ for all $i,j$. Since the sign of 
$h_{ij}$ does not change in a small neighborhood of $a$,  
$\R^k- S$ would contain some ball around $a$.  
And this contradicts the assumption that $S$ is dense 
in $\R^k$. 

Let $D$ and $L$ be the upper bounds on the degree and bit-size of
the polynomials occurring in $F'$, given by Theorem~\ref{pro:qe}.
According to Lemma~\ref{le:constr-ts}, 
we can compute a point $\alpha\in\Z^k$
satisfying $h_{ij}(\alpha)\ne 0$, for all $i,j$
and thus $F(\alpha)$, by a straight-line program with
$\Oh(k\log D + \log L)$
arithmetic operations. By plugging in the bounds on $D$ and
$L$ the claim follows
(use $(\prod_i n_i)^{1/w}\le n/w$).
\endproof

\subsection{Partial witness sequences}
\label{se:pws}

Let $K\subseteq\R$ and $\alpha\in\R^k$ with components algebraically 
independent over $K$. By Remark~\ref{re:guq}(i)(c), 
for any formula $F$ with coefficient field contained in~$K$, 
the implication $(\forall^\ast a\, F(a)) \Rightarrow F(\alpha)$
holds. Thus $\alpha$ may be interpreted as a {\em partial witness} for 
$\forall^\ast a\, F(a)$.

\begin{remark}
The converse of the implication above, i.e., 
$F(\alpha)\Rightarrow (\forall^\ast a\, F(a))$ 
does not hold in general. Actually, a point $\a\in\R^k$ as 
above and such that $F(\a)$ is true only ensures Euclidean 
genericity: we have 
$F(\alpha)\Rightarrow (\exists^*a F(a))$. 

Over $\C$, the equivalence  
$(\forall^*a F(a))\Leftrightarrow F(\a)$ holds 
since Euclidean and Zariski genericity are equivalent. 
\end{remark} 

Given a formula $F(u,a)$ we are now interested in partial 
witnesses for its Zariski-genericity property which can be used 
for all values of the parameter $u$. This may not be attainable 
with a single partial witness, but it turns out to be doable by using 
short sequences of such witnesses and taking a majority vote. 
Recall that $[n]$ denotes the set $\{1,\ldots,n\}$. 

\begin{definition}\label{def:pws}
Let $F(u,a)\in\FR$ with free variables 
$u\in\R^p$ and $a\in\R^k$. A sequence 
$\balpha=(\balpha_1,\ldots,\balpha_{2p+1})\in(\R^k)^{(2p+1)}$ 
is called a {\em partial witness sequence} for $F$ if 
\begin{equation}\label{eq:def-pws}
 \forall u\in\R^p\ \Big( 
   \forall^\ast a\in\R^k\, F(u,a) \Longrightarrow 
  |\{i\in [2p+1] \mid F(u,\balpha_i)\}| > p \Big) .
\end{equation}
We denote the set of partial witnesses of $F$ by $\PW(F)$. 
\end{definition}

\begin{lemma}\label{le:PW-dense}
$\PW(F)$ is Zariski dense in $\R^{k(2p+1)}$.
\end{lemma}

\begin{proof}
The proof is by a transcendence degree argument 
similar as in \cite[Thm.~5.1]{koir:97-1}. 
Let $K$ be the coefficient field of $F$.
We interpret~(\ref{eq:def-pws}) as a first order formula in~$\FR$ 
with free variables $\balpha_1,\ldots,\balpha_{2p+1}$ and 
coefficient field~$K$. Applying Remark~\ref{re:guq}(i)(c) to 
this formula, it is enough to show that $\balpha\in\PW(F)$ 
for any $\balpha\in\R^{k(2p+1)}$ with components  
algebraically independent over $K$. 

Take such $\balpha$ and let $u\in\R^p$.
Let $K'$ be the field extension of $K$ generated by the 
components of $u$ and let $K''$ be the field extension of 
$K'$ generated by the components of $\balpha$. Then the 
transcendence degree of $K''$ 
over $K'$ is at least $k(2p+1)-p$. Let $B$ be a transcendence 
basis of $K''$ over $K'$ consisting of components of $\balpha$. 
Then $B$ can omit components of at most $p$ of the $\balpha_i$'s. 
The remaining $\balpha_i$'s have algebraically independent 
components over $K'$ and therefore $F(u,\balpha_i)$ holds true 
for them. Thus 
$|\{i\in [2p+1] \mid F(u,\balpha_i)\}| > p$. 
\end{proof}

The next theorem is similar to~\cite[Thm.~3]{koir:99a}. 

\begin{theorem}\label{th:cons-pws}
\begin{description}
\item[(i)]
Let $F(u,a)\in\FRbit$ be in prenex form with free variables
$u\in\R^p$ and $a\in\R^k$, 
$n$ bounded variables, $w$ alternating quantifier blocks, and 
$m$ atomic predicates given by polynomials of degree at 
most $\delta\ge 2$  
with integer coefficients of bit-size at most~$\ell$. 
Then a point 
$\balpha\in \PW(F)\cap\Z^{k(2p+1)}$ can be computed 
by a division-free straight-line program $\Gamma$ of length 
$(kp)^{\Oh(1)}\, n^w \log (m\delta) + \Oh(\log\ell)$ having $1$ 
as its only constant and no inputs. 
\item[(ii)]
There exists a Turing machine which, with input 
$(p,k,n,w,m,\delta,\ell)$, computes~$\Gamma$ in time polynomial 
in the length of $\Gamma$. This machine does not depend 
on~$F$.  
\end{description}
\end{theorem}

\begin{proof}
We first replace the formula $F$ by a quantifier free formula $F'$
according to Theorem~\ref{pro:qe}. 
Let $M$ be the number of atomic predicates of $F'$, 
and $D$ and $L$ be the degree and the bit-size of the occuring 
polynomials, respectively. We have
$$
 \log D \le \Oh(n^w\log (m\delta)),\quad  \log L \le 
 \Oh(n^w\log (m\delta) + \log (p+k+\ell)),
$$
and
$$
 \log M \le \Oh(k n^w\log (m\delta)).
$$
We replace the generic quantifier in formula~(\ref{eq:def-pws})
according to Remark~\ref{re:guq}(i)(b) and  thus write the formula 
as
$$
 \forall u\ \forall \epsilon\ \forall a\ \exists a'\ 
  \Big( \epsilon \le 0 \vee \big( F'(u,a') \wedge \|a-a'\| < \epsilon \big)
  \Longrightarrow
  \bigvee_I \bigwedge_{i\in I} F'(u,\balpha_i) \Big),
$$
where $I$ runs over all $p+1$-element subsets of $[2p+1]$. 
This formula, let us call it~$\psi$, defines $\PW(F)$ and is therefore 
Zariski-generically true by Lemma~\ref{le:PW-dense}. 
We may therefore apply Proposition~\ref{pro:con-gen-pt} to 
the prenex formula $\psi$.
Note that $\psi$ has $k(2p+1)$ free variables and  
$2k+p+1 $ bounded variables, two quantifier blocks, and polynomials of 
degree at most $D$ and bit-size at most $L$.
The number of atomic predicates of $\psi$ equals $(2p+2)M+2$. 
Proposition~\ref{pro:con-gen-pt} therefore implies that 
we may compute an integer point in $\PW(F)$ by a straight-line 
program with 
$\Oh(kp (k+p)^2 \log (MD) + \log L)$ arithmetic operations. 
The latter can be bounded by 
$(kp)^{\Oh(1)}\, n^w \log (m\delta) + \Oh(\log\ell)$. This shows part~(i). 
Part~(ii) follows from part~(ii) in 
Proposition~\ref{pro:con-gen-pt}.
\end{proof}

\begin{remark}\label{rem:BSS}
\begin{description}
\item[(i)]
It follows from part (ii) Theorem~\ref{th:cons-pws} that 
the element $\balpha$ in part~(i) of this theorem can be 
computed by a machine over $\R$ or $\C$, upon 
input $(p,k,n,w,m,\delta,\ell)$, in time order of 
the length of $\Gamma$. Note, however, that this computation 
may not be possible within these time bounds in the classical 
setting since the bit-size of the components in 
$\balpha$ grows exponentially fast. 
\item[(ii)]
Over the field $\C$ one can define the stronger notion of 
{\em witness sequence}. For this we replace in formula~(\ref{eq:def-pws}) 
of Definition~\ref{def:pws} the implication from left to right 
by an equivalence. The analogue of Lemma~\ref{le:PW-dense} is 
then true and therefore witness sequences can be computed by 
``short'' straight-line programs as in Theorem~\ref{th:cons-pws}. 
This approach was taken in Koiran~\cite{koir:97-1} to devise 
a method to compute dimensions of algebraic sets in $\NPC$. 
Over the reals one cannot work with witness sequences, but the 
method can be saved by working with partial witness sequences 
as described above.
\end{description}
\end{remark}

\section{Complexity of the geometric degree}
\label{sec:degree}

The (geometric) degree $\deg Z$ of an algebraic 
variety $Z$ embedded in affine or projective space
can be interpreted as a measure for the degree of nonlinearity 
of $Z$.
A detailed treatment of this notion 
can be found in standard textbooks on algebraic 
geometry~\cite{hart:77,mumf:76,shaf:74}.  
In this section ``dimension'' always 
refers to complex dimension. 

\begin{definition}
Let $Z\subseteq\C^n$ be an algebraic set of dimension~$d\geq0$.  
If $Z$ is irreducible then its (geometric) {\em degree} $\deg Z$ 
is the number of intersection points of $Z$ with a generic affine subspace 
of codimension~$d$. If $Z$ is reducible then its 
degree is the sum of the degrees of all irreducible components of 
$Z$ of maximal dimension.\footnote{We note here that 
in algebraic complexity it is common to define the degree of 
a reducible variety as the sum of the degrees of 
{\em all} irreducible components (cf.\ \cite{bucs:96}).} 
The degree of the empty set is defined as $0$. 
\end{definition}

We are going to study the following problem in the computational 
model of machines over $\C$. 

\begin{description}
\item{$\DEGREE$} ({\em Geometric degree})\quad  
Given a finite set of complex multivariate polynomials,
compute the geometric degree of its affine zero set. 
\end{description}

Here is the main result of this section. 

\begin{theorem}\label{th:degree}
The problem $\DEGREE$ is $\FPC^{\CPCi}$-complete for Turing 
reductions.
\end{theorem}

The difficult part of the proof is the upper bound, 
i.e., the membership of $\DEGREE$ to $\FPC^{\CPCi}$. 
To show this membership, we have to 
describe a polynomial time algorithm over $\C$, 
which computes the degree using oracle calls to $\CPCi$. 
The basic idea of our $\DEGREE$ algorithm 
is very simple. Let $f_1,\ldots,f_r$ be an instance for 
$\DEGREE$ and denote its zero set by $Z$. We first compute 
the dimension $d=\dim Z$ by 
calls to $\HNC$-oracles using Theorem~\ref{th:dim}. 
By definition, $\deg Z$ is the number of intersection points 
of $Z$ with a generic affine subspace $A$ of codimension~$d$. 
If we could compute such an $A$, then the number of intersection 
points could be obtained by a call to $\#\HNC$. 

The difficulty is how to compute a generic affine subspace. 
Of course, the obvious way to turn this idea into 
an algorithm would be to choose the subspace~$A$ at random. 
This would yield a randomized algorithm for computing the degree. 
However, our goal is to choose $A$ deterministically. We will do so 
using partial witness sequences for parametrized 
formulas as described in Section~\ref{se:gen-quan}, for 
which we need to concisely express the degree. 
If $A_a$ denotes an affine subspace of $\C^{n}$ of 
codimension $d$ encoded by the parameter $a\in\C^h$, then we have  
by the definition of degree
\begin{equation}\label{eq:def-deg}
     \forall^\ast a\in\C^h\ \ |Z \cap A_a| = \deg Z.
\end{equation}
It is clear that the above statement can 
be expressed by a first-order formula over~$\C$. However, the obvious 
way to do this leads to a formula with exponentially many variables 
since $\deg Z$ can be exponentially large. 

Our goal is thus to express~(\ref{eq:def-deg}) in a more concise way. 
This will be achieved by using the notion of transversality 
(see Lemma~\ref{le:mumf}). 
However, the translation of the transversality condition into a 
concise first order formula is a little subtle and  will require 
some further ideas (see Lemma~\ref{le:F-first-order}). 

\subsection{Smoothness and transversality}
\label{se:smoothness}

An important notion in algebraic geometry is that of a smooth 
point in a variety. To define smoothness we use Zariski tangent 
spaces.  

\begin{definition}
Let $Z\subseteq\C^n$ be an algebraic set, $x\in Z$, and 
$f_1,\ldots,f_r$ be generators of the vanishing ideal~$\mI(Z)$ of~$Z$. 
The {\em Zariski tangent space} $T_xZ$ of $Z$ at $x$ is defined by 
$$
  T_xZ=\mZ(d_xf_1,\ldots,d_xf_r)
$$
where the {\em differential of $f$ at $x$}, $d_xf\colon\C^n\to\C$, is 
the linear function defined by 
$d_xf X=\sum_{j=1}^n\partial_{X_j}f(x)X_j$. 
We say that $x$ is a {\em smooth point } of $Z$ if 
the dimension of $T_xZ$ equals the local dimension $\dim_x Z$ 
of $Z$ at $x$. 
A point in $Z$ which is not smooth is said to be a 
{\em singular point} of $Z$. 
\end{definition}

\begin{remark}\label{re:rad}
Note that $T_xZ$ is easy to compute from a set of generators 
of~$\mI(Z)$, but it may not be so, if instead we only have at 
hand an arbitrary set of polynomials with zero set $Z$. 
\end{remark}

\begin{definition}
Let $Z\subseteq\C^n$ be an algebraic set of dimension~$d$
and $A\subseteq\C^n$ be an affine subspace  of codimension~$d$.
\begin{enumerate}
\item $A$ is called {\em transversal to $Z$ at $x\in Z\cap A$} iff  
$x$~is a smooth point of~$Z$ and 
$T_x Z \oplus T_x A=\C^n$. 

\item We say that $A$ is {\em transversal} to $Z$ when  
$A$ is transversal to $Z$ at all intersection points $x\in Z\cap A$
and if, additionally, there are no intersection points of $Z$ and~$A$ at infinity.
No intersection points at infinity means that 
$\overline{Z} \cap \overline{A} \subseteq \C^n$, 
where $\overline{Z}$ and $\overline{A}$ are the projective closures 
in $\proj^n(\C)$ of $Z$ and $A$. 
\end{enumerate}
\end{definition}


In the following, we will parametrize affine subspaces of 
codimension~$d$ as follows. We denote by 
$A_a\subseteq\C^n$ the affine 
subspace of $\C^{n}$ described by the system of linear 
equations $g_1(x)=0,\ldots,g_d(x)=0$ with coefficient vector 
$a\in\C^h$, where $h=d(n+1)=\Oh(n^2)$. 
Note that $\dim A \ge n-d$ for all $a$ and 
$\forall^\ast a\, \dim A_a =n-d$. 

The following lemma shows that the transversality of $A$ to $Z$ 
can be used to certify that the number of 
intersection points of $Z$ and $A$ equals $\deg Z$.

\begin{lemma}\label{le:mumf}
If $Z\subseteq\C^n$ is an algebraic set of dimension~$d$ and $h=d(n+1)$, 
then we have:
\begin{description} 
\item[(i)]
$\forall^\ast a\in\C^h$ $A_a$ is transversal to $Z$  
\item[(ii)] 
$\forall a \in\C^h\ \big(\mbox{$A_a$ is transversal to $Z$} 
\Longrightarrow |Z\cap A_a| = \deg Z \big)$. 
\end{description}
\end{lemma}

\begin{proof} 
This lemma is proved in Mumford~\cite[\S 5A]{mumf:76}
for irreducible projective varieties~$Z$. It remains to show that it 
extends to the case where~$Z$ is affine and reducible. 
Let $Z_1,\ldots,Z_t$ be the irreducible components of $Z$.
A dimension argument shows that for a generic~$a$, 
$A_a$ does neither meet the components~$Z_i$ of dimension less than~$d$, 
nor the intersections $Z_i\cap Z_j$ for $i < j$. Similarly, 
$\overline{A_a}$ does not meet $\overline{Z_i}-Z_i$ 
for generic~$a$. Hence (i) follows from the corresponding statement for 
irreducible projective varieties.

For proving~(ii) we assume that $A_a$ is transversal to $Z$. 
Then $\mathrm{codim}A_a=d$ and each point $x\in Z\cap A_a$ 
is a smooth point of $Z$ of local dimension~$d$. Hence there is 
exactly one irreducible component of $Z$ passing through~$x$ and 
this component has dimension~$d$. 
We therefore have $|Z\cap A_a| = \sum_{i=1}^s |Z_i \cap A_a|$ 
where $Z_1,\ldots,Z_s$ denote the irreducible components of dimension~$d$. 
Moreover, $A_a$ is transversal to each of these $Z_i$, hence 
$|Z_i \cap A_a|= \deg Z_i$ by \cite[\S 5A, Thm.~5.1]{mumf:76}. 
Altogether, we obtain 
$|Z\cap A_a| = \sum_{i=1}^s \deg Z_i = \deg Z$
by the definition of the degree of reducible algebraic sets.
\end{proof}

\subsection{Expressing smoothness and transversality} 
\label{se:transv}

Lemma~\ref{le:mumf} suggests to use transversality to concisely 
express degree. But, in turn, to express transversality a 
difficulty may arise. When we try to describe the Zariski 
tangent space of $Z$ at a point $x$, the given equations 
$f_1=0,\ldots,f_r=0$ for $Z$ might not generate the vanishing 
ideal of $Z$, since multiplicities might occur. 
In other words, the ideal generated by $f_1,\ldots,f_r$ 
might be different from the radical ideal,  
and it is not clear how to compute generators of the radical 
within the resources allowed. As a way 
out, we will express the tangent space and the transversality 
condition at~$x$ by a first order 
formula, in which all information regarding $Z$ is given by a unary 
predicate expressing membership of points to $Z$.\footnote{This is 
closely related to the question of the expressive power of query 
languages for constraint spatial databases~\cite{kupl:00}.}

To do so we will use the notion of intersection multiplicity, 
so we next recall some facts about it. For more on this, the book by 
Mumford~\cite{mumf:76} is an excellent reference 
fitting well our geometric viewpoint.\footnote{Mumford considers 
projective varieties, but the following 
local considerations clearly hold in the affine setting as well.} 

\begin{definition}
Assume that $Z\subseteq\C^n$ is an irreducible variety of 
dimension~$d$ and let $A_a\subseteq\C^n$ be an affine subspace 
of codimension~$d$ as above.
Suppose that $x$ is an isolated point of $Z\cap A_a$. 
Then, by~\cite[Cor.~5.3]{mumf:76}, there exists a positive 
integer~$i$ satisfying that  
for every sufficiently small Euclidean neighborhood 
$U\subseteq\C^n$ of $x$ there is a Euclidean neighborhood 
$V\subseteq\C^h$ of $a$ such that for all $a'\in V$ 
\begin{equation}\label{eq:char-mult}
\mbox{$A_{a'}$ is transversal to $Z$ $\Rightarrow$ 
      $|Z\cap A_{a'}\cap U| = i$}.
\end{equation}
We call $i$ the {\em intersection multiplicity} of $Z$ and $A_a$ at $x$ 
and we denote this number by $i(Z,A_a;x)$.  
The {\em multiplicity} $\mathrm{mult}_x(Z)$ of $Z$ at $x$ is 
defined as the minimum of $i(Z,A_a;x)$ over all affine linear 
subspaces $A_a$ of 
codimension~$d$ such that $x$ is an isolated point of $Z\cap A_a$
\cite[Def.~5.9]{mumf:76}. It is known that $x$ is a smooth 
point of $Z$ iff $\mathrm{mult}_x(Z) = 1$ \cite[Cor.~5.15]{mumf:76}.
\end{definition}

The following lemma is essential for the first order characterization
we are seeking.

\begin{lemma}\label{le:char-trans}
Let $Z\subseteq\C^n$ be an algebraic set of dimension~$d$
and $A_a\subseteq\C^n$ be an affine subspace of codimension~$d$, 
parametrized as above. For $x\in Z\cap A_a$ the 
following two conditions are equivalent:
\begin{description}
\item[(a)] $A_a$ is transversal to $Z$ at $x$. 
\item[(b)] 
For every sufficiently small Euclidean neighborhood 
$U\subseteq\C^n$ of $x$ there is a Euclidean neighborhood 
$V\subseteq\C^h$ of $a$ such that for all $a'\in V$ 
the intersection $Z\cap A_{a'}\cap U$ contains exactly one point.
\end{description}
\end{lemma}

\begin{proof}
(a) $\Rightarrow$ (b). Assume that 
$\varphi_1(x')=0,\ldots,\varphi_{n-d}(x')=0$ are local equations 
of $Z$ at $x$ (i.e., they generate the vanishing ideal of $Z$ in 
the localization at $x$). Let $g_1(a,x')=0,\ldots,g_d(a,x')=0$ 
be equations for $A_a$, parametrized by the coefficient vector 
$a\in\C^h$. The transversality of $Z$ and $A_a$ 
at $x$ implies that the Jacobian matrix at $x$ of the polynomial 
map
$$
 \C^n\to\C^n, x'\mapsto (\varphi_1(x'),\ldots,
 \varphi_{n-d}(x'),g_1(a,x'),\ldots,g_d(a,x'))
$$
is invertible. 
The implicit function theorem tells us that there 
is a continuous map $s\colon V_0\to U_0$ between
Euclidean open neighborhoods $V_0$ of $a$ and $U_0$ of $x$ such that 
for all $a'\in V_0$, $s(a')$ is the unique solution in $U_0$ of 
the system of equations
$$
 \varphi_1(x')=0,\ldots,\varphi_{n-d}(x')=0,\ g_1(a,x')=0,
 \ldots,g_d(a,x')=0 .
$$ 
For any Euclidean neighborhood $U\subseteq U_0$ of $x$, the 
Euclidean neighborhood $V:=s^{-1}(U)$ satisfies the statement of 
condition~(b). 

(b) $\Rightarrow$ (a). 
By contraposition, we assume that $A_a$ is not transversal to 
$Z$ at $x$ and show that condition (b) is not satisfied by 
considering several cases.  

Suppose first that $\dim_x Z<d$. Let $U'$ denote the open neighborhood 
of $x$ consisting of the set of points in $\C^n$, which do not lie in an 
irreducible component of $Z$ of dimension~$d$.
Then $Z\cap A_{a'}\cap U'=\emptyset$ for Zariski almost all $a'\in\C^h$.
If condition (b) were satisfied, there would exist sequences 
$x_i\to x$ and $a_i\to a$ such that $x_i\in Z\cap A_{a_i}$ and 
$Z\cap A_{a_i}\cap U'=\emptyset$ for all $i$. 
Hence $x_i\not\in U'$ for all $i$, which contradicts the fact 
that $x_i$ converges to $x$. Thus (b) is violated. 

In the following we assume that $\dim_x Z=d$. 
We may assume that $x$ is an isolated point of $Z\cap A_a$ since  
otherwise, (b) is clearly not satisfied.
We will distinguishing several cases and prove that condition~(b) 
is violated by showing the following claim in each case:
\begin{equation}\label{eq:claim}
{\begin{minipage}{12cm}
\noindent There are two sequences $(x_i)$ and $(x_i')$ in $\C^n$,  
both converging to~$x$, and there is a sequence $(a_i)$ in $\C^h$ converging to~$a$
such that $x_i,x_i'\in Z\cap A_{a_i}$ and $x_i\ne x_i'$ for all~$i$.
\end{minipage}}
\end{equation}

Let $Z_1$ be an irreducible component of $Z$ passing 
through $x$ such that $\dim Z_1 =d$.  
If $x$ is a singular point of $Z_1$, then  
$i(Z_1,A_a;x)\geq \mathrm{mult}_x(Z_1)\ge2$ 
and claim~(\ref{eq:claim}) follows by the characterization~(\ref{eq:char-mult}) 
of the multiplicity. 

We may therefore assume that $x$ is a smooth point of $Z_1$. 
If $A_a$ is not transversal to $Z_1$ at~$x$, then
$T_xZ_1 \cap T_xA \ne 0$ and therefore
$T_xA$ contains a line $\ell$ tangent to $Z_1$ at $x$. 
There is a sequence of points $x_i\in Z_1$, $x_i\ne x$, 
converging to $x$ in the Euclidean topology such that 
the secant $s_i$ through $x$ and $x_i$ converges to $\ell$. 
Take $A_{a_i}$ to be the affine space of codimension $d$ 
spanned by $\ell^{\perp}$ and $s_i$ 
(here $\ell^{\perp}$ is the orthogonal 
complement of $\ell$ in $A_a$). 
Since $A_{a_i}\cap Z\supseteq\{x,x_i\}$,  
and we can achieve that $a_i\to a$ for a suitable 
choice of the parameter $a_i$, 
the claim~(\ref{eq:claim}) follows. 

We are left with the case where $A_a$ is transversal to $Z_1$ at~$x$.
Since $A_a$ is not transversal to $Z$ at~$x$, there must be at least
one further irreducible component $Z_2$ of $Z$ passing through~$x$. 
Consider a sequence of points $x_i\in Z_2 - Z_1$ converging to $x$ and 
such that $x_i\ne x$. Consider also points $z_1,\ldots,z_{n-d}$ 
in $A_a$ such that the vectors $z_1-x,\ldots,z_{n-d}-x$ are 
linearly independent. Now let $A_{a_i}$ be the affine space 
of codimension $d$ passing through $x_i,z_1,\ldots,z_{n-d}$. 
We can achieve that $a_i\to a$.

On the other hand, since $A_a$ is transversal to $Z_1$ at~$x$, 
we may apply condition~(b) to $Z_1$ and $A_a$.
Passing over to a subsequence of $(A_{a_i})$, we 
obtain that there is a sequence $x_i'\in Z_1\cap A_{a_{i}}$ 
converging to $x$. This shows the claim~(\ref{eq:claim})
and completes the proof of the lemma.
\end{proof}

In the following, we parametrize a system $f_1,\ldots,f_r$ 
of polynomials over $\C$ by its vector of non-zero 
coefficients $u\in\C^q$, and we denote the corresponding 
zero set by~$Z_u$. (Hence we use the sparse encoding, 
cf.~\S\ref{se:salg}.) 
Recall that we parametrize affine subspaces $A_a\subseteq\C^n$ 
of codimension~$d$ by elements $a\in\C^h$.

\begin{lemma}\label{le:F-first-order}
For all $0\le d\le n$ 
there is a first order formula $F_d(u,a)$ in $\FRbit$
in prenex form with seven quantifier blocks,  
$\Oh(n^2)$ bounded variables, and with $\Oh(q+n)$ 
atomic predicates given by integer polynomials 
of degree at most $\delta$ and bit-size $\Oh(1)$, 
such that for all $u\in\C^q\simeq\R^{2q}$ with 
$\dim_{\C} Z_u = d$ and all $a\in\C^h$: 
$$
 \mbox{$F_d(u,a)$ is true} \Longleftrightarrow 
 \mbox{$A_a$ is transversal to $Z_u$.}
$$
\end{lemma}

\begin{proof} 
In what follows, we interpret all occuring formulas over $\C$ as 
first order formulas in $\FR$ by encoding a complex 
number by its real and imaginary part.  


Suppose that $A_a$ is of codimension~$d$. Then 
property (b) in Lemma~\ref{le:char-trans} expressing 
transversality of $A_a$ to $Z_u$ at~$x$ can be written as the 
following formula $\varphi(u,a,x)$: 
\begin{equation*}
\begin{split}
\exists\, \epsilon_0 >0\ &\forall\, 0 < \epsilon < \epsilon_0\ 
\exists\, \delta>0\ 
  \forall a'\in\C^h\ \exists y\in\C^n\ \forall\, z\in\C^n\
  \Big( \|a-a'\|<\delta\ \wedge\ \\
 & \|y-x\|<\epsilon\ \wedge\ \|z-x\|<\epsilon\ 
 \wedge\ y\in Z_u\cap A_{a'}\ \wedge\  \left(z\in Z_u\cap A_{a'} 
    \Longrightarrow y=z\right)\Big).
\end{split}
\end{equation*}
The property that $A_a$ is transversal to $Z_u$ at all affine 
intersection points $x\in Z_u\cap A_a$ then reads as:
$$
 \forall x\in\C^n\ \big( x\in Z_u\cap A_a \Longrightarrow 
 \varphi(u,a,x)). 
$$
The property that $Z_u$ and $A_a$ have no intersection points at 
infinity is expressed by
$$
 \forall x\in\C^{n+1} \big( x\in\overline{Z}_u \wedge 
    x\in\overline{A}_a 
     \Longrightarrow x_0\ne 0 \big),
$$
where the bar denotes projective closure 
(we have now an additional homogenizing variable $x_0$). 
We express the predicate $x\in\overline{Z}_u$ in the form 
$$
 \forall \epsilon > 0\ \exists x'\in\C^{n}\ 
 \exists \lambda\in \C-\{0\}\   
 (x'\in Z_u \wedge \|x-\lambda(1,x')\| < \epsilon ) ,
$$
using the fact that the Zariski-closure of constructible sets 
equals the Euclidean closure. 

Finally, we can express that $\mathrm{codim} A_a = d$ 
by requiring that there exists a linear subspace $L$ 
with $\dim L \ge d$ and $A_a^{\rm lin} \cap L = 0$, where  
$A_a^{\rm lin}$ denotes the linear space associated with 
$A_a$. 

Altogether, we see that the transversality condition can be expressed 
by a formula in $\mathcal{F}_{\R}$ of the required description size.
\end{proof}

\begin{remark}
\begin{description}
\item[(i)]
It is not clear whether transversality can be expressed by 
short first order formulas over $\C$ since the Euclidean topology 
is involved. 
We will circumvent this difficulty by working with the first order 
theory over the reals. The next lemma provides a concise 
first order (over the reals) characterization of transversality.
However, it is important to keep in mind 
that we will resort to the reals only as a way of reasoning. 
All computations in the proof of Theorem~\ref{th:degree} will 
be done by machines over $\C$.

\item[(ii)] Note that the projective closure $\overline{Z_u}$ is 
included in but may not be equal to 
the zero set of the homogenization of the polynomials defining $Z_u$. 
\end{description}
\end{remark}

\subsection{Proof of Theorem~\ref{th:degree}}

We begin with the membership of $\DEGREE$ to $\FPC^{\CPCi}$.  
Let $p=2q$. Then, 
by Theorem~\ref{th:cons-pws} and Remark~\ref{rem:BSS}(i), 
a partial witness sequence 
$\balpha=(\balpha_1,\ldots,\balpha_{2p+1})$ for 
the formula $F_d(u,a)$ in Lemma~\ref{le:F-first-order} 
can be computed by a machine over $\C$, 
given input $(p,k,n,w,m,\delta,\ell)$, in time 
$(nq)^{\Oh(1)}\,\log\delta$.   
Note that this quantity is polynomially bounded in the sparse 
input size $\Oh(nq\log\delta)$. 

We claim the correctness of the following algorithm for $\DEGREE$. 

\algorithm
{\bf input} $f_1,\ldots,f_r$ with coefficient vector~$u$\\
compute $d:=\dim Z_u$ by oracle calls to $\HNC$ using 
Theorem~\ref{th:dim}\\ 
compute a partial witness sequence 
$\balpha=(\balpha_1,\ldots,\balpha_{2p+1})$ of $F_d(u,a)$\\ 
{\bf for} $i=1$ {\bf to} $2p+1$\\ 
\>   compute $N_i := |Z_u \cap A_{\balpha_i}|$ by an oracle call  
     to $\#\HNC$\\ 
compute the majority $N$ of the numbers $N_1,\ldots,N_{2p+1}$\\
{\bf return} $N$
\falgorithm 
Put $I:=\{i\in [2p+1]\mid F_d(u,\balpha_i)\mbox{ holds}\}$. 
Lemma~\ref{le:F-first-order} and Part~(ii) of 
Lemma~\ref{le:mumf} imply that $N_i=\deg Z_u$ 
for all $i\in I$.  
Part~(i) of Lemma~\ref{le:mumf} tells us that 
$\forall^\ast a\, F_d(u,a)$. 
Since $\balpha$ is a partial witness sequence, this implies that 
$|I| > p$ (cf.\ Definition~(\ref{def:pws})). 
This proves the claim.

It is obvious that the above algorithm can be implemented as a polynomial time 
oracle Turing machine over $\C$. This shows the membership. 

To prove the hardness, note that, 
by Theorem~\ref{completeness}, $\#\HNC$ is $\CPCi$-complete.   
It is therefore sufficient to Turing reduce 
$\#\HNC$ to $\DEGREE$. The following reduction does so. 
For a given system of equations first decide whether its 
solution set $Z$ is zero-dimensional by a call to $\HNC$ 
using Theorem~\ref{th:dim}. This call to $\HNC$ can be replaced by 
a call to $\DEGREE$ since $\HNC$ reduces to $\DEGREE$ 
(recall $Z=\emptyset$ iff $\deg Z= 0$).  
If $\dim Z=0$, then compute $N:=\deg Z$ by a call to $\DEGREE$ and 
return~$N$, otherwise return $\infty$. 
\hfill$\Box$

\section{Preliminaries from algebraic and differential topology}
\label{se:ADT}

\subsection{Euler characteristic of compact semialgebraic sets}
\label{se:EUC}

It is well known that any compact semialgebraic set~$S$ 
can be triangulated~\cite[\S9.2]{bocr:87}.  
Instead of working with triangulations, we will use 
the more general notion of finite cell complexes, 
since this is necessary for 
the application of Morse theory in \S\ref{se:morse}. 
Compact semialgebraic sets are homeomorphic to finite cell 
complexes and their topology can be studied through 
the combinatorics of cell complexes. 

We briefly recall the definition of a finite cell complex 
(also called finite CW-complex), 
see, for instance,~\cite{hatc:02} for more details.   
We denote by $D^n$ the closed unit ball in $\R^n$, and by 
$S^{n-1}=\partial D^n$ its boundary, the $(n-1)$-dimensional 
unit sphere. An {\em $n$-disk} is a space homemorphic to $D^n$. 
By an {\em open $n$-cell} we understand a space $e^n$ homeomorphic 
to the open unit ball $D^n - \partial D^n$.  
A (finite) {\em cell complex} $X$ is obtained by the 
following inductive procedure.

We start with a finite discrete set $X^0$, whose points are 
regarded as $0$-cells. 
Inductively, we form the {\em $n$-skeleton} $X^n$ from $X^{n-1}$ by 
attaching a finite number of open $n$-cells $e^n_\alpha$ via 
continuous maps $\varphi_\alpha\colon S^{n-1}\to X^{n-1}$. 
This means that $X^n$ is the quotient space of the disjoint union 
$X^{n-1}\sqcup_\alpha D_\alpha^n$ of $X^{n-1}$ with a finite 
collection of $n$-disks $D_\alpha^n$ under the identifications 
$x\equiv \varphi_\alpha(x)$ for 
$x\in\partial D_\alpha^n=S^{n-1}$. Thus as a set, 
$X^n=X^{n-1}\sqcup_\alpha e_\alpha^n$, where each $e_\alpha^n$ is 
an open $n$-cell. We stop this procedure after finitely many 
steps obtaining the compact space $X=X^d$ of dimension $d$. 

We note that each cell $e^n_\alpha$ has a {\em characteristic map} 
$\Phi_\alpha\colon D^n_\alpha\to X$ which extends the attaching map 
$\varphi_\alpha$ and is a homeomorphism from the interior of 
$D^n_\alpha$ onto $e^n_\alpha$. Namely, we can take $\Phi_\alpha$ 
to be the composition 
$D^n_\alpha\hookrightarrow X^{n-1}\sqcup_\alpha D_\alpha^n
  \to X^n\hookrightarrow X$,
where the middle map is the quotient map defining $X^n$.  

\begin{example}\label{ex:1} 
\begin{description}
\item[(i)]
The $n$-sphere can be realized as a cell complex 
with two cells, of dimension $0$ and $n$, respectively. 
The cell $e^n$ is attached to $e^0$ by the constant map 
$\varphi: S^{n-1}\to e^0$. 
\item[(ii)]
Real projective space $\proj^n(\R)$ is defined as the space 
of all lines through the origin in $\R^{n+1}$. This is 
equivalent to identify antipodal points in $S^n\subset\R^{n+1}$, 
a presentation which in addition yields a natural topology 
in $\proj^n(\R)$ ---the quotient topology induced by the 
identification. Removing the southern hemisphere, this is 
yet equivalent to the space obtained by keeping the northern 
hemisphere and identifying antipodal points in the equator. 
Since the northern hemisphere (without the equator) is 
homeomorphic to $e^n$ and the equator with identified  
antipodal points is just $\proj^{n-1}(\R)$, it follows that 
$\proj^{n}(\R)$ is obtained from the $n+1$ cells 
$e^0,e^1,\ldots,e^n$ by taking $X_0=e^0$ and, inductively,
obtaining $X_k=\proj^k(\R)$ from $X_{k-1}$ by attaching 
$e^k$ via the identification of antipodal points 
$\varphi_k:\partial D^k\to X^{k-1}$. 
\item[(iii)]
Complex projective space $\proj^n(\C)$ (already seen 
in~\S\ref{se:salg}) is the quotient of the unit sphere 
$S^{2n+1}\subset\C^{n+1}$ for the equivalence relation 
$v\equiv \lambda v$ for all $\lambda\in\C$ with 
$|\lambda|=1$. A reasoning as the one above (taking into 
account that equivalence classes are now homeomorphic to~$S^1$) 
shows that $\proj^n(\C)$ is obtained from the $n+1$ cells 
$e^0,e^2,\ldots,e^{2n}$ as above, now getting 
$X_{2k}=\proj^k(\C)$, for $k=0,\ldots,n$.  
\end{description}  
\end{example}

The {\em Euler characteristic} of a cell complex $X$ is defined as 
$\chi(X) = \sum_{k=0}^d (-1)^k N_k$, where $N_k$ is the number 
of $k$-cells of the complex. It is a well-known fact that 
$\chi(X)$ depends only on the topological space $X$ and not on the 
cellular decomposition. That is, if two cell complexes are 
homeomorphic, then their Euler characteristics are the same. 
Actually $\chi$ is even a homotopy invariant.

\addtocounter{example}{-1}
\begin{example}{\bf (continued)}\quad 
For the spaces considered above we obtain, using their 
cell decompositions, that 
$$
 \chi(S^n)=\left\{\begin{array}{ll}
           2&\mbox{if $n$ is even}\\
           0&\mbox{if $n$ is odd}\end{array}\right.
 \qquad\qquad\qquad
 \chi(\proj^n(\R))=\left\{\begin{array}{ll}
           1&\mbox{if $n$ is even}\\
           0&\mbox{if $n$ is odd}\end{array}\right.
$$
and $\chi(\proj^n(\C))=n+1$. 
\end{example}

%

A continuous map $p\colon X\to Y$ between topological spaces 
is called a {\em covering map} if there exists an open cover $\{U_\alpha\}$  
of $Y$ such that for each $\alpha$, $p^{-1}(U_\alpha)$ is a disjoint 
union of open sets in $X$, each of which is mapped by $p$ 
homeomorphically onto $U_\alpha$ (see e.g., \cite[III.3]{bred:93}). 
If the cardinality of the fibre $p^{-1}(y)$ is 
constant for $y\in Y$, then this cardinality is called the 
{\em number of sheets} of the covering map. This condition is 
satisfied when $Y$ is connected.

\begin{lemma}\label{le:efb}
If $X\to Y$ is a covering map with $m$ sheets ($m$ finite) and $Y$ is 
a cell complex, then $X$ is also a cell complex and $\chi(X)=m\chi(Y)$. 
\end{lemma}

\begin{proof}
The characteristic maps $D^n\to Y$ lift to $X$ in exactly $m$ ways. 
This way, one obtains a cell decomposition of $X$ with the number 
of $k$-cells exactly $m$ times the number of $k$-cells of $Y$. 
Thus the alternating sum of these numbers for~$X$ is $m$ times 
the alternating sum of these numbers for $Y$.
For more details see \cite[Prop.~13.5, p.~216]{bred:93}.
\end{proof}

\subsection{Non-compact semialgebraic sets}

There are several ways to extend the definition of $\chi$ to 
non-compact sets. The usual one uses singular homology 
and preserves the property of $\chi$ of being homotopy 
invariant. In \S\ref{se:modif} we will see another way which 
does not, but instead has a useful additivity property. 

In algebraic topology one assigns to a topological space $X$ 
and a field $F$ the singular {\em homology vector spaces} $H_k(X;F)$ 
for $k\in\N$, which depend only on the homotopy type of $X$ 
and $F$. 
The {\em $k$th Betti number over $F$} $b_k(X;F)$ of $X$ is defined 
as the dimension of $H_k(X;F)$. In case $F=\Q$ we write 
$b_k(X)$ and talk about the $k$th Betti number of $X$.  
The Euler characteristic of the space $X$ is defined by  
\begin{equation}\label{eq:def-ec}
 \chi(X)=\sum_{k\in\N} (-1)^k \dim_F H_k(X;F) 
\end{equation}
(if this sum is finite). 
The Betti numbers $b_k(X;F)$ depend 
on the field $F$ as well as on $X$. Remarkably, their alternate 
sum, is independent of $F$. In addition, for cell complexes $X$, 
this alternate sum coincides with $\chi(X)$ as defined 
in~\S\ref{se:EUC}. For a general reference to homology we refer 
to~\cite{hatc:02,munk:84}. 

More generally, one can assign to a pair $Y\subseteq X$ of 
topological spaces the {\em relative Euler characteristic} 
$\chi(X,Y):=\chi(X)-\chi(Y)$. 
It can also be characterized in terms of the 
{\em relative homology vector spaces} $H_k(X,Y;F)$ as 
$\chi(X,Y)=\sum_{k\in\N} (-1)^k \dim_F H_k(X,Y;F)$.
Since $H_k(X,Y;F)$ depends only on the homotopy type of the 
pair $(X,Y)$, the same holds for the relative Euler characteristic 
$\chi(X,Y)$. Note that $H_k(X,\emptyset;F)=H_k(X;F)$ and 
$\chi(X,\emptyset)=\chi(X)$.  

\begin{lemma}\label{le:ec}
Let $Z$ be a compact real algebraic $n$-dimensional manifold and 
$K\subseteq Z$ be a compact semialgebraic subset. Then
$$
 \chi(Z-K)=\left\{\begin{array}{ll} 
   \chi(Z) -\chi(K) & \mbox{ if $n$ is even,}\\ 
   \chi(K) & \mbox{ if $n$ is odd.}
        \end{array}\right. 
$$
\end{lemma}

\begin{proof}
A fundamental duality principle going back to Poincar\'e 
and extended by 
Alexander and Lefschetz states that 
for an $n$-dimensional manifold $Z$ and a compact subset $K$
carrying the structure of a cell complex, the relative homology space 
$H_k(Z,Z-K;\Z_2)$ is isomorphic to the homology 
space\footnote{Actually, one gets a natural isomorphism with 
the cohomology vector space $H^{n-k}(X;\Z_2)$ induced by the 
$\Z_2$-orientation of the manifold $Z$, but this is not important 
for our purposes.} $H_{n-k}(K;\Z_2)$ 
for all $k$. See~\cite[Prop.~3.46, p.~256]{hatc:02} 
or~\cite[Thm.~8.3, p.~351]{bred:93}.
Therefore, we get under the assumptions of the theorem that 
\begin{eqnarray*}
 \chi(Z)-\chi(Z-K) &=&\chi(Z,Z-K) 
          =\sum_k (-1)^k\dim H_k(Z,Z-K;\Z_2) \\ 
         &=& (-1)^n\sum_k (-1)^k \dim H_k(K;\Z_2) = (-1)^n \chi(K) .
\end{eqnarray*}
This implies the claim in the case where $n$ is even. 
When $n$ is odd, we obtain that $\chi(Z-K) = \chi(K) + \chi(Z)$. 
On the other hand, applying the above formula for $K=Z$ yields 
$\chi(Z) = -\chi(Z)$ and thus $\chi(Z)=0$. 
Hence $\chi(Z-K)=\chi(K)$. 
\end{proof}

\subsection{Modified Euler characteristic} 
\label{se:modif}

Let $S$ be the disjoint union of two semialgebraic sets 
$S_1$ and $S_2$. In general, it is not true that 
$\chi(S)=\chi(S_1)+\chi(S_2)$ . For a counterexample, 
consider the closed $3$-dimensional unit ball $D^3$ 
decomposed into its interior $e^3$ and its boundary $S^2$.   

Yao~\cite{ayao:92} defined the {\em modified Euler characteristic} $\mchi$  
of semi-algebraic sets, which satisfies an additivity property,
and coincides with the usual Euler characteristic 
for compact semialgebraic sets. The following proposition 
from~\cite{ayao:92} characterizes this notion.

\begin{proposition}\label{pro:mod-euler}
There is a unique function $\mchi$ 
mapping semialgebraic sets to integers, which  
satisfies the following properties:
\begin{description}
\item[(i)] If $S=\bigsqcup_{i=1}^N S_i$ is a disjoint union 
of semialgebraic sets, 
then $\mchi(S)=\sum_{i=1}^N \mchi(S)$. 

\item[(ii)] We have $\mchi(S)=\chi(S)$ for compact semialgebraic sets.

\item[(iii)] If there is a semialgebraic homeomorphism $S\to T$, then 
$\mchi(S)=\mchi(T)$.  
\end{description}
\end{proposition}

\begin{proof} 
For the proof of existence, which  relies on 
Hironaka's triangulation theorem~\cite{hiro:75} 
for bounded (not necessarily closed) semialgebraic sets, 
we refer to~\cite{ayao:92}. 

The proof of uniqueness shows that, in principle, the computation of
$\mchi$ can be reduced to computations of $\chi$ for compact
semialgebraic sets.  Since this is useful for calculating some
examples, and to familiarize the reader with the notion of the 
modified Euler characteristic, we present the simple proof of 
uniqueness.

Any unbounded semialgebraic set
$S\subseteq\R^n$ is semialgebraically homeomorphic to a bounded one. 
Namely, $S$ is homeomorphic to its its image  
under the inverse of the stereographic projection 
$S^n - \{(0,\ldots,0,1)\} \to\R^n, x\mapsto y$ 
given by the equations $y_i =x_i/(1-x_{n+1})$.
Therefore, by property~(iii), it suffices to show uniqueness for 
bounded semialgebraic sets~$S$. We proceed by induction on the 
dimension of~$S$. The case $\dim S\le 0$ is clear.  
Consider the disjoint union $\overline{S} = S \cup R$, where 
$R:=\overline{S}-S$. We have 
$\dim R \le \dim\partial S < \dim S$ since $R$ is contained 
in the boundary $\partial S$ of~$S$, cf.~\cite[Prop.~2.8.12]{bocr:87}.
Since $\overline{S}$ is compact we have 
$\mchi( \overline{S})=\chi(S)$ by property~(ii). 
Property~(i) implies that 
$\mchi(S) = \chi(\overline{S}) - \mchi(R)$, 
hence $\mchi(S)$ is determined by $\mchi(R)$, which in turn is uniquely 
determined by the induction hypothesis. 
\end{proof}

\begin{example} 
The inverse image of $\R^n$ under the stereographic projection is 
$S^n$ minus a point, hence $\mchi(\R^n)=\chi(S^n) - 1 = (-1)^n$. 
Note that, in contrast with $\chi$, $\mchi$ is not invariant 
under homotopies.  
\end{example}

\begin{corollary}\label{cor:add}
If $S_1,\ldots,S_N$ are semialgebraic subsets of $\R^n$,  then 
we have 
$$
 \mchi\big(\bigcup_{i=1}^N S_i\big) = 
 \sum_{I\ne\emptyset} (-1)^{|I|-1}\mchi\big(\bigcap_{i\in I} S_i\big),
$$
where the summation is over all nonempty subsets $I$ of~$[N]$. 
\end{corollary}

\begin{proof}
This follows from the inclusion-exclusion principle 
taking into account that $\mchi$ behaves additively 
with respect to disjoint unions.
\end{proof}

\subsection{Locally closed spaces and Borel-Moore homology}

A noncompact locally closed set $S$ can be 
compactified by adding just one point.
More specifically, there is a compact semi-algebraic set $\dot{S}$  
and a continuous semi-algebraic map $\iota\colon S\to\dot{S}$, 
which is a homeomorphism onto its image, such that $\dot{S}-\iota(S)$ 
consists of just one point~$\infty$, cf.~\cite[2.5.9]{bocr:87}.

Let $S$ be a locally closed semialgebraic set and $F$ be a field.
If $S$ is not compact, then the {\em Borel-Moore homology vector spaces} 
of~$S$ over $F$ are defined as the relative homology spaces of the pair 
$(\dot{S},\infty)$, that is,  
$H_k^\bm(S;F):=H_k(\dot{S},\infty;F)$, cf.~\cite[\S11.4]{bocr:87}.
If $S$ is compact, then we define $H_k^\bm(S;F)=H_k(S,F)$.
Moreover, we define the $k$th {\em Borel-Moore Betti number} of $S$, 
denoted $b_k^\bm(S)$, as the dimension of $H_k^\bm(S;\Q)$.
Thus we have $b_k^\bm(S)=b_k(S)$ for compact~$S$. 

From the above, the following well-known characterization easily 
follows.

\begin{proposition}
Let $S$ be a locally closed semialgebraic set. Then 
$$
 \mchi(S)= \sum_{k\in\N} (-1)^k b_k^\bm(S).
$$
\end{proposition}

\begin{proof}
If $S$ is compact the result is trivial. Otherwise, 
we have $\mchi(S)=\mchi(\dot{S})-\mchi(\infty) = 
\chi(\dot{S})-\chi(\infty) = \chi(\dot{S},\infty)$.
On the other hand
$$
 \chi(\dot{S},\infty)= \sum_{k\in\N}(-1)^k\dim H_k(\dot{S},\infty;\Q) 
 = \sum_{k\in\N}(-1)^k\dim H_k^\bm(S;\Q),
$$
which shows the assertion. 
\end{proof}

\begin{remark}\label{re:1ptcomp} 
\begin{description}
\item[(i)] $\mchi(S)$ can also be interpreted as 
the Euler characteristic of $S$ with respect to the cohomology 
$H^*_{\rm c}(S;\Q)$ of $S$ with compact supports, a notion naturally 
occuring in the Poincar\'e duality theorem for noncompact 
manifolds, cf.~\cite[\S 3.3, p.~242]{hatc:02}.  

\item[(ii)] 
It is an important fact that for a complex algebraic 
variety~$W$ we have $\mchi(W)=\chi(W)$. 
If $W$ is smooth of complex dimension $n$, then this follows 
from the Poincar\'e duality 
$H_k(W)\simeq H^{2n-k}_{\rm c}(W)$, using the 
interpretation of $\mchi(W)$ as the Euler characteristic of 
the cohomology $H^*_{\rm c}(W)$ with compact support. For the proof 
of the general case see~\cite[Exercise~\S4.5, p.~95 and 
Notes~\S4.13, p.~141]{fult:93}. 
\end{description}
\end{remark}  

\subsection{Morse Theory}
\label{se:morse}

We recall now some notions and facts from Morse theory. A 
general reference for this is~\cite{miln:63}. 

Let $Z$ be a differentiable manifold and $\varphi\colon Z\to\R$ 
be differentiable. A point $x\in Z$ is a {\em critical point} of 
$\varphi$ if the differential 
$d_x\varphi\colon T_xZ \to \R$ vanishes. In this case, 
one may consider the {\em Hessian} 
$H_x\varphi\colon T_x Z \times T_x Z \to \R$
of $\varphi$ at $x$, which is a symmetric bilinear form
(defined by the second order derivatives of $\varphi$ in 
local coordinates). The function $\varphi$ is called 
{\em nondegenerate} at the critical point $x$ 
if its Hessian is nondegenerate at $x$. 
The function $\varphi$ is called a {\em Morse function} if 
all its critical points are nondegenerate.

We call the number of negative  
eigenvalues of a symmetric 
matrix or of a symmetric bilinear form its {\em index}. 
The {\em index} of $\varphi$ at $x$ is defined as the index of 
$H_x\varphi$. Throughout the paper, we will use the convenient 
notation $\{\varphi\le r\}:=\{x\in Z\mid \varphi(x)\le r\}$.

The main theorem of Morse theory~\cite[Thm.~3.5]{miln:63} 
states the following.  

\begin{theorem}\label{th:morse}
Assume that $\varphi\colon Z\to\R$ is a Morse function 
on a differentiable manifold $Z$ with finitely many 
critical points. Moreover, assume that 
$\{\varphi\le r\}$ is compact for all $r\in\R$. 
Then $Z$ has the homotopy type of 
a cell complex with one cell of dimension~$k$ for each 
critical point of $\varphi$ of index~$k$.  
\hfill$\Box$
\end{theorem}

We will use the following consequence of this result, 
adapted to the semialgebraic setting. 

\begin{corollary}\label{pro:morse}
Let $Z\subseteq\R^n$ be a real algebraic manifold. Then,  
\begin{description}
\item[(i)] 
The Euclidean distance function $L_a\colon Z\to\R$, 
$x\mapsto \|x-a\|^2$, is a Morse function for Zariski 
almost all $a\in\R^n$. 
\item[(ii)] 
Suppose that $L_a$ is a Morse function on $Z$. Then the number 
$N_k$ of critical points of $L_a$ with index $k$ is finite 
for all $0\le k\le n$ and $\sum_{k=0}^n (-1)^k N_k$ 
equals the Euler characteristic $\chi(Z)$ of $Z$. 
\end{description}
\end{corollary}

\begin{proof}
{\bf (i)\quad} 
The first claim follows as in~\cite[\S 6]{miln:63} by using the 
semialgebraic Morse-Sard Theorem~\cite[Thm.~9.5.2]{bocr:87}. 

{\bf (ii)\quad}  
It is easy to see that the set of critical points of $L_a$ is 
semialgebraic. Moreover, critical points are isolated. Since 
semialgebraic sets have finitely many components, it follows 
that there are only finitely many critical points. 
Note that $Z\cap\{x\in\R^n\mid L_a(x)\le r\}$ is compact for 
all $r\in\R$. Hence we can apply Theorem~\ref{th:morse} 
and the claim follows from the definition of $\chi$. 
\end{proof}

Let $\mH$ be the set of polynomials 
$f\in\R[X_1,\ldots,X_n]$ satisfying that $\mZ(f)\neq\emptyset$ along 
with the regularity condition
\begin{equation}\label{eq:hs-reg}
\forall x\in\R^n\ (f(x)=0 \Rightarrow \grad f(x) \ne 0 ).
\end{equation}
Note that $\mZ(f)$ is a smooth hypersurface for $f\in\mH$. 

Consider $f\in\mH$ and $Z=\mZ(f)$. Then $x\in Z$ is a critical 
point of $L_a$ if and only if $\sum_k\partial_{X_k}f(x) (x_k-a_k) =0$. 
Let $x$ be a critical point of $L_a$ such that (w.l.o.g.) 
$\partial_{X_n}f(x)\ne 0$. By the implicit function theorem, 
locally around $x$, $Z$ is the graph of a function 
$(t_1,\ldots,t_{n-1})\mapsto y(t_1,\ldots,t_{n-1})$ 
which defines a local coordinate system around $x$,  
$$
  (t_1,\ldots,t_{n-1})\mapsto 
  x(t) := (t_1,\ldots,t_{n-1}, y(t_1,\ldots,t_{n-1})).
$$

\begin{lemma}\label{le:Hessian}
The Hessian $(H_{ij})= (\partial_{t_i}\partial_{t_j} L_a(x(t)))$ 
of the distance function $L_a$ at $x$ in terms of the local coordinates $t_i$ is given by 
\begin{eqnarray*}
\lefteqn{  \frac{1}{2}(\partial_{X_n}f)^2 H_{ij}=}\\ 
 & &(\partial_{X_n}f)^2 \delta_{ij} + 
  (\partial_{X_i} f)(\partial_{X_j} f) + (x_n-a_n) 
  ((\partial_{X_i} f)(\partial_{X_j}\partial_{X_n} f) 
  - (\partial_{X_i}\partial_{X_j} f)(\partial_{X_n}f) ).
\end{eqnarray*}
\end{lemma}

\begin{proof}
By differentiating $L_a(x)=\sum_k (x_k-a_k)^2$ with respect 
to $t_i$ we obtain 
$$
 \partial_{t_i} L_a = 2 \sum_{k=1}^n (x_k -a_k)\, \partial_{t_i} x_k =
  2(t_i - a_i + (y-a_n)\partial_{t_i} y) .
$$
Differentiating again with respect to $t_j$ yields
$$
 H_{ij} =\partial_{t_i}\partial_{t_j} L_a = 
 2 (\delta_{ij} + (\partial_{t_i} y)(\partial_{t_j} y) 
   + (y -a_n)\, \partial_{t_i}\partial_{t_j} y).
$$
From $f(t_1,\ldots,t_{n-1}, y(t_1,\ldots,t_{n-1}))=0$  we get 
$\partial_{t_i} y = -\frac{\partial_{X_i}f}{\partial_{X_n}f}$ 
by differentiating. Differentiating this again with respect to 
$t_j$ we obtain
$$
 \partial_{t_i}\partial_{t_j} y = 
 \frac{-(\partial_{t_j}\partial_{X_i} f)(\partial_{X_n}f) 
  +(\partial_{X_i} f)(\partial_{t_j}\partial_{X_n}f)}
   {(\partial_{X_n}f)^2}.
$$
By plugging these expressions for the partial derivatives of 
$y$ into the above formula for $H_{ij}$ and taking into account 
that $t_i=x_i$ for $i<n$ we obtain the asserted formula. 
\end{proof}

As in Section~\ref{sec:degree}, we denote by $u\in\R^p$ 
the vector of non-zero coefficients of the polyomial $f=f_u$ of 
degree~$\delta$ in $X_1,\ldots,X_n$, and 
write $Z_u:=\mZ(f_u)$ for its zero set in $\R^n$. 

The following lemma gives a certificate for $L_a$ to be a 
Morse function on $Z_u$ in form of a parametrized first 
order formula. It plays a similar role for the completeness 
proof of $\MEULER$ as the certificate for transversality 
for the completeness proof of $\DEGREE$, which was provided in  
Lemma~\ref{le:F-first-order}.

\begin{lemma}\label{le:gen-morse}
There is a first order formula $F(u,a)$ in $\FRbit$ 
in prenex form with one quantifier block, 
$n$ bounded variables, and with $\Oh(n)$ 
atomic predicates given by integer polynomials 
of degree at most $\Oh(n\delta)$ and bit-size $\Oh(n\log (np))$ 
such that, for all $u\in \R^p$ such that $f_u\in\mH$ 
and all $a\in\R^n$, the following holds: 
$$
 \mbox{$F(u,a)$ is true} \Longleftrightarrow 
  \mbox{$L_a\colon Z_u\to\R$ is a Morse function.} 
$$
\end{lemma}

\begin{proof}
The fact that $L_a\colon Z_u\to\R$ is a Morse function can be 
expressed by the following formula
{\small 
$$
 \forall x\in\R^n\ \bigg(f(x)=0\,\wedge\, 
  \sum_{k=1}^n\partial_{X_k}f(x) (x_k-a_k) =0 \Longrightarrow 
  \bigvee_{k=1}^n \big( \partial_{X_k} f(x) \ne 0 \wedge \det 
   H_xL_a \ne 0
   \big)\bigg) 
$$ 
}
where, we recall, $H_xL_a$ denotes the Hessian of $L_a$ at $x$. 
We now replace $H_xL_a$ by the explicit expression for it 
given in Lemma~\ref{le:Hessian}, 
after making the appropriate changes due to the fact that we require 
the $k$th partial derivative of $f$ to be nonvanishing at~$x$ 
instead of the $n$th derivative. The assertion follows now easily 
by inspecting the above formula.
\end{proof}

\section{Complexity of the Euler characteristic}
\label{sec:EulerReal}

Another main result of this paper proves the 
completeness in $\FPR^{\CPRi}$ of the following problem over $\R$. 

\begin{description}
\item{$\MEULER$} ({\em Modified Euler characteristic})\quad 
Given a semialgebraic set  
$S\subseteq\R^n$ as a union of basic semialgebraic sets
$$
\mbox{$S=\bigcup_{i=1}^t\{x\in\R^n\mid g_i(x)=0, 
   f_{i1}(x)>0,\ldots,f_{ir_i}(x)>0\}$},
$$ 
decide whether $S$ is empty and if not, compute $\mchi(S)$. 
\end{description} 


\begin{theorem}\label{th:euler_sa}
The problem $\MEULER$ is $\FPR^{\CPRi}$-complete with 
respect to Turing reductions. 
\end{theorem}

The upper bound in Theorem~\ref{th:euler_sa} is proved in 
several steps: in Section~\ref{se:bs-p-a-var} we reduce the 
basic semialgebraic case to the case of a smooth hypersurface. 
This case is then treated in 
Section~\ref{se:smrhs} based on Morse theory and the concept of 
partial witness sequence developed in Section~\ref{se:pws}.  
Finally, we combine these two ingredients in 
Section~\ref{se:ass} 
to treat the case of arbitrary semialgebraic sets,
using the inclusion-exclusion principle, which is possible 
due to the additivity property of the modified Euler characteristic.

\subsection{Basic semialgebraic, projective and affine varieties}
\label{se:bs-p-a-var}

\begin{lemma}\label{le:pa-euler} 
Let $g,f_1,\ldots,f_r\in\R[X_1,\ldots,X_n]$ be of degree at most~$\delta$
and $S:=\{x\in\R^n\mid g(x)=0, f_{1}(x)>0,\ldots,f_{r}(x)>0\}$. 
Put $g_0:=g$ and define for $1\le i\le r$
$$ 
 g_i:=X_{n+i}^2 f_i - 1,\ G_i:=X_0^{\delta+3}g_i(X_1/X_0,\ldots,X_{n+r}/X_0),
 \  H:=\sum_{i=0}^r G_i^2.   
$$
Then, $\Phi:=\mZ(H-1)\subset\R^{n+r+1}$ is a smooth affine 
hypersurface and 
$$
 \mchi(S) = \frac{(-1)^{n+r}}{2^{r+1}} (2 -\chi(\Phi)). 
$$
\end{lemma}

\begin{proof}
Note that, for $i=1,\ldots,r$, 
$G_i\in\R[X_0,\ldots,X_{n+r}]$ is homogeneous and 
$g_i\in\R[X_1,\ldots,X_{n+r}]$.
Define the affine variety $Y_a$ and the projective variety $Y_p$ by 
$$
 Y_a:=\mZ(g_0,\ldots,g_r)\subseteq\R^{n+r},\ 
 Y_p:=\mZ(G_0,\ldots,G_r)=\mZ(H)\subseteq\proj^{n+r}(\R). 
$$
For $\epsilon\in\{-1,1\}^r$ consider the open subsets 
$Y_\epsilon := Y_a \cap \left(\cap_{i=1}^r \{\mathrm{sgn}(x_{n+i}) 
=\epsilon_i \}\right)$ of~$Y_a$.
Clearly, each $Y_\epsilon$ is semialgebraically homeomorphic to $S$. 
Moreover, $Y_a$ is the disjoint union of the $Y_\epsilon$. 
Hence 
\begin{equation}\label{eq:E1}
 2^r \mchi(S) = \sum_\epsilon\mchi(Y_\epsilon) = \mchi(Y_a) . 
\end{equation}

Consider the open subset $V:=Y_p\cap \{X_0\ne 0\}$ of $Y_p$, 
which is semialgebraically homeomorphic to~$Y_a$.
Since we homogenized with exponent $\delta+3$, which is one higher 
than the maximum degree $\delta+2$ of the $g_i$, we have 
$Y_p - V = \mZ_{\proj^{n+r}(\R)}(X_0)\simeq \proj^{n+r-1}(\R)$. 
By additivity of $\mchi$ we have 
$\chi(Y_p)=\chi(\proj^{n+r-1}(\R)) + \mchi(V)$, hence 
\begin{equation}\label{eq:E2}
 \mchi(Y_a)=\mchi(V)=
   \left\{\begin{array}{ll} 
   \chi(Y_p) & \mbox{ if $n+r$ is even,}\\ 
   \chi(Y_p) -1  & \mbox{ if $n+r$ is odd.}
   \end{array}\right. 
\end{equation}

Note that $1$ is a regular value of $H$, since $H=(\deg H)\sum_i X_i \partial_{X_i} H$ 
by the homogeneity of $H$. Hence the ``Milnor fibre''
$$
   \Phi:= \{x\in\R^{n+r+1} \mid H(x)=1\}
$$
is a smooth affine hypersurface. 
Put $U:= \{x\in\proj^{n+r}(\R)\mid H(x)\ne 0\}$. 
We claim that the canonical map
$$
 \pi\colon \Phi\to U,\ (x_0,\ldots,x_{n+r}) \mapsto (x_0:\cdots:x_{n+r})
$$
is a covering map with two sheets. Indeed, 
$\pi^{-1}(U\cap \{X_i\ne 0\}) = (\Phi \cap \{X_i >0\}) \cup (\Phi\cap \{X_i<0\})$, 
and $\pi$ induces homeomorphisms from both $\Phi \cap \{X_i > 0\}$ and $\Phi\cap \{X_i<0\}$ 
to $U\cap \{X_i\ne 0\}$, respectively. 

By Lemma~\ref{le:efb} we have $\chi(\Phi)=2\chi(U)$. 
On the other hand, by Lemma~\ref{le:ec} and Example~\ref{ex:1}, we get 
$\chi(U)=1 -\chi(Y_p)$ if $n+r$ is even and 
$\chi(U)=\chi(Y_p)$ if $n+r$ is odd.
Altogether, we obtain
\begin{equation}\label{eq:E3}
 \chi(Y_p)=\left\{\begin{array}{ll} 
   1 -\frac{1}{2}\chi(\Phi) & \mbox{ if $n+r$ is even,}\\ 
   \frac{1}{2}\chi(\Phi) & \mbox{ if $n+r$ is odd.}
   \end{array}\right. 
\end{equation}
Combining Equations~(\ref{eq:E1}), (\ref{eq:E2}), 
and~(\ref{eq:E3}) the assertion follows.
\end{proof}

\subsection{The case of a smooth real hypersurface}
\label{se:smrhs}

Consider the function 
$\chi_{\mH}:\mH\to\Z,\ f\mapsto\chi(\mZ(f))$
computing the Euler characteristic of the smooth hypersurface 
$\mZ(f)$ given by $f\in\mH$. Note that we don't consider the 
modified Euler characteristic here.

\begin{proposition}\label{pro:seuler}
The function $\chi_{\mH}$ belongs to $\FPR^{\CPRi}$. 
\end{proposition}

\begin{proof}
Let $\INDEX$ be the following decision problem.
An input to $\INDEX$ is a tuple $(u,a,x,k,J)$, 
where $u$ encodes a real polynomial $f$ in $n$ variables,  
$a,x\in\R^n$, $k\in\N$ and $J\subseteq [n]$ is nonempty. 
The question is to decide whether $x$ is a critical point of 
index~$k$ of the function $L_a\colon Z_u\to\R$ 
satisfying $\partial_{X_j}f(x)\ne 0$ for all $j\in J$. 

The problem $\INDEX$ is in $\PR$. Indeed, given the tuple 
$(u,a,x,k,J)$, one first computes the Hessian 
$H_xL_a$ by using Lemma~\ref{le:Hessian}. Then, 
one computes its characteristic polynomial   
(a computation known to be in $\FPR$, see~\cite{bcss:95,bucs:96},  
and finally one uses Sturm's algorithm to compute the number of 
real zeros in the interval $(-\infty,0)$ (again in 
$\FPR$, see~\cite{gage:99}). Comparing this number with $k$ decides 
$\INDEX$ for $(u,a,x,k,J)$. 

Given $(u,a)$, let $\chi_+(u,a)$ denote the number 
of $(x,k,J)$ such that $(u,a,x,k,J)\in\INDEX$ 
and $k+|J|$ is odd. 
Similarly, we define $\chi_-(u,a)$ by requiring that $k+|J|$ is 
even. Since $\INDEX\in\PR$, the functions 
$\Ri\x\Ri\to\N\cup\{\infty\}$ mapping $(u,a)$ to $\chi_+(u,a)$ 
and $\chi_-(u,a)$, respectively, are in $\CPRi$. 


We claim that
\begin{equation}\label{eq:chi+-}
 \mbox{$\chi(Z_u) = \chi_+(u,a) - \chi_-(u,a)$ if $L_a$ 
       is a Morse function on $Z_u$.} 
\end{equation}
In fact, if $N_k$ denotes the number of critical points of 
$L_a$ on $Z_u$, we have $\chi(Z_u)=\sum_k (-1)^k N_k$ by 
Corollary~\ref{pro:morse}. Let $N_{kJ}$ denote the number of 
critical points~$x$ of $L_a$ on $Z_u$ satisfying 
$\partial_{X_j}f(x)\ne 0$ for all $j\in J$. 
By the inclusion-exclusion principle we have 
$N_k = \sum_{J\ne\emptyset} (-1)^{|J|-1} N_{kJ}$
(note that at every point of $Z_u$ at least one partial derivative 
does not vanish due to the regularity condition for $f_u$).  
This implies 
$$
 \chi(Z_u)=\sum_k (-1)^k N_k
  =\sum_{k,J\ne\emptyset} (-1)^{k+|J|-1} N_{kJ} 
  = \chi_+(u,a) - \chi_-(u,a)
$$
as claimed in (\ref{eq:chi+-}). 

Lemma~\ref{le:gen-morse} and Theorem~\ref{th:cons-pws} imply that a 
partial witness sequence $\alpha$ for 
the first order formula $F(u,a)$ certifying that 
$L_a\colon Z_u\to\R$ is a Morse function 
can be computed (uniformly) by a division-free straight-line 
program with $(np)^{\Oh(1)}\,\log(\delta)$  
arithmetic operations, using $1$ as the only constant. 

The following algorithm computing $\chi_{\mH}$ can be 
implemented as a polynomial time 
oracle Turing machine querying oracles in $\CPRi$. 
\algorithm
{\bf input} $f\in\mH$ encoded by its coefficient vector~$u$\\
compute a partial witness sequence 
 $\balpha=(\balpha_1,\ldots,\balpha_{2p+1})$ of $F(u,a)$\\ 
{\bf for} $\ell=1$ {\bf to} $2p+1$\\ 
\>\>compute $\chi(u,\balpha_\ell) := \chi_+(u,\balpha_\ell) - \chi_-(u,\balpha_\ell)$ \\ 
compute the majority $\chi(u)$ of the numbers 
 $\chi(u,\balpha_1),\ldots,\chi(u,\balpha_{2p+1})$\\
{\bf return} $\chi(u)$
\falgorithm 

In order to show that this algorithm actually computes the Euler 
characteristic of its input, put 
$\Lambda:=\{\ell\in [2p+1]\mid F(u,\balpha_\ell)\mbox{ holds}\}$. 
By definition of $F$ we know that $L_{\balpha_\ell}$ is a Morse 
function on $Z_u$ for all $\ell\in \Lambda$. Hence, 
by (\ref{eq:chi+-}), $\chi(Z_u)=\chi(u,\balpha_\ell)$ for 
all $\ell\in \Lambda$. 
On the other hand, by Proposition~\ref{pro:morse}(i) we have 
$\forall^\ast a\, F(u,a)$. 
Since $\balpha$ is a partial witness sequence, this implies that 
$|\Lambda| > p$ (cf.\ Definition~(\ref{def:pws})). 
Therefore, the algorithm indeed computes the Euler 
characteristic of $Z_u$.
\end{proof}

\subsection{Arbitrary semialgebraic sets}
\label{se:ass}

\begin{proposition}\label{pro:euler_sa}
The problem $\MEULER$ is contained in $\FPR^{\CPRi}$.
\end{proposition}

\begin{proof} 
Consider an instance $S=\cup_{i=1}^t S_i$ of the problem $\MEULER$, 
where $t\ge 1$ and 
$S_i=\{x\in\R^n\mid g_i(x)=0, f_{i1}(x)>0,\ldots,f_{ir_i}(x)>0\}$.  
Emptyness of $S$ can be easily decided in $\FPR^{\CPRi}$. 

Assume $S\neq\emptyset$. By adding dummy inequalities $1>0$, 
we may assume that $r_i=r$ for all~$i$. 
Corollary~\ref{cor:add} tells us that 
\begin{equation}\label{eq:chi1}
 \mbox{$\mchi(S) = \sum_{I\ne\emptyset} (-1)^{|I|-1}\mchi (S_I)$}, 
\end{equation}
where for $I\subseteq [t]$, the basic semialgebraic set 
$S_I\subseteq\R^n$ is defined by
$$
S_I:= \bigcap_{i\in I} S_i  =
    \bigg\{x\in\R^n\mid \sum_{i\in I} g_i(x)^2 =0,\ f_{ij}(x) > 0 
          \mbox{ for $i\in I, j\in [r]$}\bigg\}.
$$
We will assume that each $S_I$ is described by exactly $rt$ 
inequalities, which can be achieved by adding further dummy 
inequalities 

According to Lemma~\ref{le:pa-euler}, we can assign to each 
nonempty index set $I\subseteq [t]$ a homogeneous polynomial 
$H_I\in\R[X_0,\ldots,X_{n+rt}]$, such that $\mchi(S_I)$ can be 
expressed by the Euler characteristic of the smooth affine hypersurface 
$\Phi_I:=\mZ(H_I-1)$ in $\R^{n+1+rt}$ as follows
\begin{equation}\label{eq:chi2}
 \mchi(S_I) = \frac{(-1)^{n+rt}}{2^{rt+1}} (2 -\chi(\Phi_I)) . 
\end{equation}
Plugging (\ref{eq:chi2}) into (\ref{eq:chi1}) 
and using that $\sum_{I} (-1)^{|I|}=0$ we obtain
\begin{equation}\label{eq:sf}
 \mchi(S) = \frac{(-1)^{n+rt}}{2^{rt+1}} 
  \bigg(2 + \sum_{I\ne\emptyset} (-1)^{|I|}\chi (\Phi_I)\bigg) .
\end{equation}

We proceed now similarly as in the proof of 
Proposition~\ref{pro:seuler}. Let $p$ be the 
number of real parameters of all the polynomials $g_i,f_{ij}$ 
involved in the above description of the set $S$. 
To emphasize the dependence on $u$, we will write
$\Phi_{I,u}$ instead of $\Phi_I$. 
For a projection point $a\in\R^{n+1+rt}$ and a parameter $u\in\R^p$ 
we consider the distance function 
$L_a\colon\Phi_{I,u}\to\R, x\mapsto \|x-a\|^2$. 

Similarly as in the proof of Proposition~\ref{pro:seuler}, we 
assign to $u\in\R^p, a\in\R^{n+1+rt}$, and $I\subseteq [t]$ 
two values $\chi_{+,I}(u,a),\chi_{-,I}(u,a)\in\N$ 
such that (cf.~(\ref{eq:chi+-}))
\begin{equation}\label{eq:mofo}
 \mbox{$\chi(\Phi_{I,u}) = \chi_{+,I}(u,a) - \chi_{-,I}(u,a)$ 
        if $L_a$ is a Morse function on $\Phi_{I,u}$.} 
\end{equation}
Namely, $\chi_{+,I}(u,a)$ is defined as the number of 
triples $(x,k,J)$, 
where $x\in\R^{n+1+rt}$, $k\in\N$, $\emptyset\ne J\subseteq [n+1+rt]$
such that $x$ is a critical point of index $k$ 
of the function $L_a\colon\Phi_{I,u}\to\R$ satisfying 
$\partial_{X_j}f(x)\ne 0$ for all $j\in J$ and such that 
$k+|J|$ is odd. 
Similarly, one defines $\chi_{-,I}(u,a)$ by requiring that 
$k+|J|$ is even. 
As in the proof of Proposition~\ref{pro:seuler}, 
one shows that the functions 
$\{0,1\}^\infty\x\Ri\x\Ri\to\N\cup\{\infty\}$ 
mapping $(I,u,a)$ to $\chi_{+,I}(u,a)$ and 
$\chi_{-,I}(u,a)$, respectively, are in $\CPRi$. 

Assume now that $u,a$ are chosen such that 
$L_a$ is a Morse function on $\Phi_{I,u}$ 
for all nonempty subsets $I$ of $[t]$. 
Plugging (\ref{eq:mofo}) into (\ref{eq:sf}) we obtain
$$
 (-1)^{n+rt}\, 2^{rt+1}\, \mchi(S)
    = 2 + \sum_{I\ne\emptyset} (-1)^{|I|}\big(\chi_{+,I}(u,a) 
      - \chi_{-,I}(u,a)\big)
    = 2 + \chi_+(u,a) - \chi_-(u,a) ,
$$
where we have put
$$
 \chi_+(u,a):=\sum_{I\ne\emptyset,\mbox{\scriptsize $|I|$ even}} 
      \chi_{+,I}(u,a),\quad  
 \chi_-(u,a):=\sum_{I\ne\emptyset,\mbox{\scriptsize $|I|$ odd}} 
      \chi_{-,I}(u,a) .
$$ 
According to Lemma~\ref{le:addition}, the 
functions $(u,a)\mapsto\chi_{+}(u,a)$ and 
$(u,a)\mapsto\chi_{-}(u,a)$ are in $\CPRi$. 

Consider the first order formula $G_I(u,a)$ in $\FRbit$ provided by  
Lemma~\ref{le:gen-morse}, which expresses the fact that 
$L_a\colon \Phi_{I,u}\to\R$ is a Morse function. 
Define the first order formula
$G(u,a):=\wedge_{I\ne\emptyset} G_I(u,a)$, which  
certifies that, for all nonempty index sets $I\subseteq [t]$,  
$L_a\colon \Phi_{I,u}\to\R$ is a Morse function. 
Theorem~\ref{th:cons-pws} and Remark~\ref{rem:BSS} 
imply that a partial witness  
sequence $\balpha=(\balpha_1,\ldots,\balpha_{2p+1})$ for 
the formula $G(u,a)$ can be computed 
in time polynomial in the input size of $S$. 
(Note that it does not harm that the number of atomic 
predicates of $G(u,a)$ is exponential in the input size of~$S$.)

After all these preparations, we see that the modified Euler
characteristic of~$S$ can be computed by essentially the same
algorithm as in the proof of Proposition~\ref{pro:seuler}. 
The modifications are as follows: 
replace the formula $F$ by $G$, reinterpret
$\chi_{+}(u,a),\chi_{+}(u,a)$ in the above way, and 
return $(-1)^{n+rt}2^{-rt-1}(2+\chi(u))$ where, again, 
$\chi(u)$ is obtained by taking a majority vote 
on the $\chi_+(u,\balpha_i)-\chi_-(u,\balpha_i)$.    
This algorithm can be
implemented as a polynomial time oracle Turing machine accessing
oracles in $\CPRi$.  The proof of correctness is identical as for the
proof of Proposition~\ref{pro:seuler}.
\end{proof}

\proofof{Theorem~\ref{th:euler_sa}}
The membership of $\MEULER$ to $\FPR^{\CPRi}$ is the 
content of Proposition~\ref{pro:euler_sa}. 
By Theorem~\ref{completeness}, $\#\FEASR$ is $\CPRi$-complete.   
To prove the Turing-hardness of $\MEULER$ for $\CPRi$, 
it is therefore sufficient to Turing reduce 
$\#\FEASR$ to $\MEULER$. The following reduction does so. 
For a given real polynomial first decide whether its 
solution set $Z$ is zero-dimensional by a call to $\FEASR$ 
using Theorem~\ref{th:dim}. This call to $\FEASR$ can be replaced by 
a call to $\MEULER$ since $\FEASR$ reduces to $\MEULER$ 
(this follows from the case distinction in the definition of 
the problem $\MEULER$).  
If $\dim Z=0$, then compute $N:=\mchi(Z)$ by a call to $\MEULER$ and 
return~$N$, otherwise return $\infty$. 
\endproof

\begin{remark}
In the papers~\cite{bruc:90,szaf:86}, the Euler 
characteristic of a real algebraic variety is expressed 
by the index of an associated gradient vector field at zero, 
which can be algebraically computed according to~\cite{eile:77}. 
Although Morse theory is not explicitly mentioned 
in~\cite{bruc:90,szaf:86}, the main idea behind these papers 
is an application of this theory as exposed in~\cite{miln:68}. 
The single exponential time algorithm in~\cite{basu:99} 
for computing the Euler characteristic uses 
Morse theory explicitly and in a crucial way. However, we note 
that the reduction in~\cite{basu:99} from the case of an arbitrary 
semialgebraic set to the case of a smooth hypersurface,  
as well as the reductions in~\cite{bruc:90,szaf:86}, 
cannot be used in our context, since it is not clear how to 
compute the deformation parameter or the sufficiently small 
radius of the intersecting sphere within the allowed resources
(polynomial time for {\em real} machines). 
Instead, we have expressed the Euler characteristic of a real 
projective variety by the Euler characteristic of its complement, 
which in turn can be expressed as the Euler characteristic 
of a ``Milnor fibre'', which is a smooth hypersurface. 
\end{remark}

\section{Completeness results in the Turing model}
\label{se:Turing}

It is common to restrict the input 
polynomials in the problems considered so far to 
polynomials with integer coefficients. The resulting problems 
can be encoded in a finite alphabet 
and studied in the classical Turing setting.
In general, if $L$ denotes a problem defined over $\R$ or $\C$, we denote
its restriction to integer inputs by $L^0$. 
This way, the discrete problems 
$\HNCbit$, $\DIMCbit$, $\DEGREEbit$, $\MEULERbit$, etc.\  
are well defined.

We are going to show next that all the above problems are 
(Turing-) complete in certain discrete complexity classes. 
These classes are obtained from real or complex complexity classes 
by the operation of taking the Boolean part. 

\subsection{Basic complete problems in Boolean parts}
\label{subsec:BP}

A problem that has attracted much attention in 
real (or complex) complexity is the computation of 
Boolean parts 
\cite{buer:97-1,cugr:97,ckklw:95,cuko:95,koir:94,koir:97a}. 
Roughly speaking, this amounts 
to characterize, in terms of classical complexity classes, 
the power of resource bounded machines over $\R$ or $\C$ 
when their inputs are restricted to be binary. 

\begin{definition}
Let ${\cal C}$ be a complexity class of decision problems 
over $\R$ or $\C$. 
Its {\em Boolean part} is the classical complexity class 
$$
  \BP({\cal C}):=\{S\cap\{0,1\}^\infty \mid 
   S\in{\cal C}\}.
$$
\end{definition}

The study of Boolean parts has been successful 
in the setting of additive machines, where practically all 
natural complexity classes have had their Boolean parts  
characterized~\cite{bucu:02,cuko:95,koir:94}. In 
contrast, much less is known in the setting of unrestricted 
machines. Two of the most significant results state that    
$\BP(\PC)\subseteq \P^{\RP}$~\cite{ckklw:95} 
and $\BP(\PAR)=\PSPACE/\poly$~\cite{cugr:97}, 
and a third one is discussed in 
Proposition~\ref{prop:compBP1} below. For stating it, 
recall that $\RP$ denotes the classical complexity class 
of problems decidable by randomized machines in polynomial 
time with (one-sided) error.  
It is well-known that 
$\P^{\RP}\subseteq\Pi^2$, 
where $\Pi^2$ denotes a class in the second 
level of the polynomial hierarchy 
(see~\cite{badg:88,papa:94} for details). 

The following upper bound for $\HNC^0$ was obtained 
by Koiran~\cite{koir:96}. 

\begin{theorem}\label{th:FEASC-bit}
$\HNC^0$ belongs to $\RP^{\NP}$ (and therefore to 
$\Pi^2$) under the generalized Riemann hypothesis GRH. 
\hfill$\Box$
\end{theorem}

A natural restriction for real or complex machines (considered 
e.g.\ in~\cite{cuko:95,koir:94,koir:97a}) is the requirement that no 
constants other than $0$ and $1$ appear in the machine program. 
Complexity classes 
arising by considering such constant-free machines are indicated 
by a superscript $0$ as in $\PR^0$, $\NPR^0$, etc. Note that we have 
already used a superscript $0$ for a decision or counting 
problem $L$ over $\R$ or $\C$, denoting by $L^0$ its 
corresponding problem restricted to integer inputs. 
Thus, the superscript $0$ means restriction to bits in two different 
contexts: machine constants when considering complexity 
classes, and input data when considering problems. This should 
not create any confusion. 

Theorem~\ref{th:FEASC-bit} provides an upper bound for 
$\HNC^0$. On the other hand, the clear $\NP$-hardness 
of $\HNC^0$ provides a lower bound. Yet there is a gap 
between $\NP$ and $\RP^{\NP}$ and the problem of how 
to close it (with regard to $\HNC^0$) is, as of today, 
an open question. 
The following result elaborates on that question.

\begin{proposition}\label{prop:compBP1}
\begin{description}
\item[(i)]
$\HNC^0$ and $\DIMC^0$ are 
$\BP(\NPC^0)$-complete. 
\item[(ii)]
Assuming GRH, we have $\NP\subseteq\BP(\NPC^0)\subseteq \RP^{\NP}$.
\end{description}
\end{proposition}  

\begin{proof}
The completeness of $\HNC^0$ in part~(i) follows from the 
following fact. The $\FPC$-reduction from an arbitrary 
$\NPC$-problem to $\HNC$ exhibited in~\cite{bcss:95}, when 
applied to a problem $L$ in $\NPC^0$, yields a 
$\FP$-reduction from $L^0$ to $\HNC^0$. This shows that 
$\HNC^0$ is $\BP(\NPC^0)$-complete. 
The completeness of $\DIMC^0$ follows from 
Theorem~\ref{th:dim}(i). 

For the reasoning above to hold it is essential that we only 
consider problems defined by $\NPC$-machines that do not use 
complex constants. Otherwise, these constants would appear as 
coefficients in the constructed polynomial system. 

The second inclusion in part~(ii) follows from part~(i) 
and Theorem~\ref{th:FEASC-bit}. The first inclusion 
is trivial.
\end{proof}

It is believed~\cite[p.~255]{papa:94}  
that $\RP^{\NP}$ has no complete problems. 
Thus, it follows from Proposition~\ref{prop:compBP1} that 
the equality $\BP(\NPC^0)=\RP^{\NP}$ is unlikely to hold. 

The rest of this section is devoted to completeness 
results in Boolean parts in the spirit of 
Proposition~\ref{prop:compBP1}. Before stating our 
result, we note that the definition of the Boolean part can be 
extended to classes such as $\CPCi$ or $\CPRi$ in an obvious way. 
Thus, we define the class of {\em geometric counting complex problems} as 
$\GCC:= \BP(\CPCi^0)$ and the class of {\em geometric counting real problems} 
$\GCR:=\BP(\CPRi^0)$.
These are classes of discrete counting problems, closed under 
parsimonius reductions, which can be located in a small region in 
the general landscape of classical complexity classes. 
Namely, we have 
$$
    \CP\subseteq\GCC\subseteq\GCR\subseteq\FPSPACE, 
$$
where the rightmost inclusion follows from 
Theorem~\ref{th:inclusions} and~\cite{cugr:97}. 

\begin{proposition}\label{prop:compBP2}
\begin{description}
\item[(i)]
$\FEASRbit$, $\SAS^0$, and $\DIMRbit$ 
are $\BP(\NPR^0)$-Turing-complete.
\item[(ii)]
$\#\SAS^0$ and $\#\FEASRbit$ are $\GCR$-Turing-complete. 
\item[(iii)]
$\#\HNC^0$ is $\GCC$-Turing-complete.
\end{description}
\end{proposition}

\begin{proof}
For the hardness 
in part (i) we use the argument in the proof 
of Proposition~\ref{prop:compBP1}(i), namely, that the reductions  
from an arbitrary $\NPR$-problem to $\FEASR$ or $\SAS$ 
yield reductions from problems in $\BP(\NPR^0)$ to 
$\FEASR^0$ or $\SAS^0$, respectively.
For the hardness in parts~(ii) and (iii) one uses the reductions in 
the proof of Theorem~\ref{completeness}.
The memberships in all statements are clear except  
for $\DIMRbit$, for which the claim follows from 
Theorem~\ref{th:dim}(ii).
\end{proof}

\begin{remark}\label{re:elimc}
One can show that 
$\BP(\NPC^0)=\BP(\NPC)$ and 
$\GCC=\BP(\CPCi^0)=\BP(\CPCi)$. 
Hence it is immaterial whether we allow the use of 
complex machine constants in the definition of 
these classes or not. Moreover, it is possible 
to extend Proposition~\ref{prop:compBP1}(ii) to
$\BP(\FPC^{\NPC})\subseteq \RP^{\NP}$, 
assuming GRH. 
The proof relies on the possibility to 
eliminate complex constants using 
witness sequences, as developed in
\cite{bcss:96,koir:97,koir:99}.
Details will be given elsewhere. 
\end{remark}

We can give some evidence that counting over $\C$ 
is indeed harder than deciding feasibility over $\C$.

\begin{corollary}\label{cor:transf}
If $\CPCi\subseteq\FPC^{\NPC}$, then the classical polynomial 
hierarchy collapses at the second level, assuming GRH.
\end{corollary}

\begin{proof}
Assuming $\CPCi\subseteq\FPC^{\NPC}$ and 
taking Boolean parts, we get by Remark~\ref{re:elimc}
$$
\CP \subseteq \BP(\CPCi) \subseteq \BP(\FPC^{\NPC})
  \subseteq \RP^{\NP} \subseteq \Pi^2.
$$
Toda's theorem ~\cite{toda:91} 
states that $\PH\subseteq\P^{\CP}$. 
Hence we conclude $\PH=\Pi^2$, which means that the 
polynomial hierachy collapses at the second level.
\end{proof}

\subsection{Degree and Euler characteristic in the Turing model}

We can now easily deduce completeness results for the 
discrete versions of the problems to compute the degree 
or the modified Euler characteristic. 

\begin{theorem}\label{th:degree-bin}
\begin{description}
\item[(i)] $\DEGREE^0$ is $\FP^{\GCC}$-complete with respect 
to Turing reductions.

\item[(ii)] $\MEULERbit$ is $\FP^{\GCR}$-complete with respect to 
Turing reductions.
\end{description}
\end{theorem}

\begin{proof}
(i) The proof given in Section~\ref{sec:degree}  
for the membership of $\DEGREE$ to $\FPC^{\CPCi}$ applies 
in our case with only one modification. 
The algorithm in the proof of Theorem~\ref{th:degree} computes 
the partial witness sequence $\balpha$ (this is done in $\FPC$) 
and then performs $2p+1$ oracle calls to $\CPCi$ to obtain the 
numbers $N_i$ for $i\in [2p+1]$. While it is clear that 
the computation of $\balpha$ is in $\BP(\FPC)$, it is equally clear 
that it is not in $\FP$ due to the exponential coefficient growth 
caused by repeated powering (cf.\ Lemma~\ref{le:constr-ts}).  
A way to solve this is  
to ``move'' the computation of $\balpha$ to the query. That is, one 
considers the problem of computing $N_i$ with input 
$(u,i)$. Clearly, this problem is in $\BP(\CPCi)$: one first 
computes $\balpha$ in $\FPC$ and then $N_i$ in $\CPCi$.    

The hardness of $\DEGREE^0$ follows as in Theorem~\ref{th:degree}
using the second statement in Theorem~\ref{th:dim}(i) instead 
of the first.  

(ii) The proof for $\MEULERbit$ is a modification of the proof of 
Theorem~\ref{th:euler_sa}, similar as for part~(i). 
\end{proof}

\begin{remark}\label{rem:BP1}
\begin{description}
\item[(i)] 
The algorithms for $\DEGREE^0$ and $\MEULERbit$ above can be further 
simplified. Since we can bound the description size of 
the formula $F(u,a)$ or $G(u,a)$ by taking into account a 
bound on the bit-size of the components of the given 
$u\in\Z^p$, the input vector~$u$ does not need to be 
considered as a parameter any more. Therefore, we may take $p=0$. 
The partial witness sequence then consists of 
a single vector $\alpha\in\Z^k$ and only one oracle call to 
$\#\HNC^0$ (or two oracle calls to $\#\FEASRbit$) are needed. 
 
\item[(ii)] 
Alternatively, the algorithm in the proof of Theorem~\ref{th:degree} 
(or Theorem~\ref{th:euler_sa}) could be modified as follows.
By part~(i) we may assume that $p=0$. 
The straight-line computation for the partial witness  
$\alpha\in\Z^{k}$ of $F$  
cannot be executed in the bit model because of the exponential 
coefficient growth.  
However, we can easily remedy this by describing the construction 
of the partial witness sequence by existentially quantifying 
over additional variables~$\beta_1,\ldots,\beta_q$ along the 
recursive description in Lemma~\ref{le:constr-ts}. 
We then query $\#\HNC^0$ for the 
system of equations in the variables $x$, $\alpha_i$ and 
$\beta_1,\ldots,\beta_q$ expressing the recursive 
construction of $\alpha_i$ and the fact that 
$x\in Z_u\cap L_{\alpha}$.
\end{description}
\end{remark}

In the Turing model we can also prove a completeness result for the 
computation of the (non-modified) Euler characteristic: 
consider the problem
\begin{description}
\item{$\EULER$} ({\em Euler characteristic for basic semialgebraic sets})\quad 
Given a basic semialgebraic set 
$S= \{x\in\R^n\mid g(x)=0, f_{1}(x)>0,\ldots,f_{r}(x)>0\}$,
decide whether $S$ is empty and if not, compute $\chi(S)$. 
\end{description}


\begin{theorem}\label{th:euler-bin}
$\EULERbit$ is $\FP^{\GCR}$-complete with respect to Turing 
reductions.  
\end{theorem}

To prepare for the proof, recall that a closed 
semialgebraic set~$S\subseteq\R^n$ has a conic 
structure at infinity~\cite[Prop. 5.50]{bapr:03}, which 
implies that there exists $r>0$ such that 
for all $r'\ge r$ there is a 
semialgebraic deformation retraction from $S$ to 
$S_{r'}:= S \cap \{x\in\R^n\mid \|x\|\le r'\}$. 
We will call $r$ a {\em cone radius  of $S$ at infinity}. 
Clearly, we have 
$\chi(S)=\chi(S_r) = \mchi (S_r)$. 

\begin{lemma}\label{le:bd-cora}
Let $p\in\Z[X_1,\ldots,X_n]$ be of degree 
at most $\delta$ with coefficients of bit-size at most $\ell$. 
Then, there exist $m=(n\delta\ell)^{\Oh(1)}$ such that  
$2^{2^m}$ is a cone radius of $\mZ(p)$ at infinity. 
\end{lemma}

\begin{proof}(Sketch)
In~\cite{geer:01} it is shown that there is a first order formula 
$\Phi(r)$ in $\FRbit$ in prenex form with the free variable $r$ such 
that there exists $r_0>0$ with 
$$
  [r_0,\infty[ \subseteq \{r\in\R\mid\mbox{$\Phi(r)$ true}\}
  \subseteq\{r\in\R\mid 
  \mbox{$r$ is a cone radius  of $\mZ(f)$ at infinity}\}. 
$$
By an inspection of the constructions in~\cite{geer:01,rann:98} 
one can show that the formula $\Phi(r)$ has a bounded number of 
quantifier blocks, $n^{\Oh(1)}$ bounded variables, and $m$ atomic 
predicates given by integer polynomials 
of degree at most~$d$ and bit-size at most $\ell'$ such that 
$
  \log (dm\ell') \le (n\delta\ell)^{\Oh(1)} .
$
The tedious details of verifying this statement about the 
desription size of $\Phi(r)$ are omitted for lack of space and 
left to the reader. 

According to Theorem~\ref{pro:qe}, the formula $\neg\Phi(r)$ 
is equivalent to a 
quantifier-free formula in disjunctive normal form 
$\vee_{i=1}^I \wedge_{j=1}^{J_i} (h_{ij}(r)\Delta_{ij}0)$, 
containing integer polynomials $h_{ij}(r)$ of 
bit-size at most~$L$ such that 
$\log L \le (n\delta\ell)^{\Oh(1)}$.

Let $\rho\in\R$ be the maximum of the real roots of the nonzero 
$h_{ij}$. We have $\rho \le 1 + \|h\|_\infty \le 1 + 2^L$. 
Note that the sign of 
$h_{ij}(x)$ is constant for $x>\rho$. Therefore, since the set 
$\{r>0\mid \neg\Phi(r)\}$ is bounded, we have
$\{r>0\mid \neg\Phi(r)\}\subseteq ]0,\rho]$.
Hence $2 + 2^L$ is a cone radius of $\mZ(f)$ at infinity, 
which proves the claim. 
\end{proof}

\proofof{Theorem~\ref{th:euler-bin}}
The hardness of $\EULERbit$ follows as in the proof of 
Theorem~\ref{th:euler_sa}. 
We prove now that $\EULERbit$ belongs to $\FP^{\GCR}$.
For given 
$S= \{x\in\R^n\mid g(x)=0, f_{1}(x)>0,\ldots,f_{r}(x)>0\}$, 
compute the polynomial
$$
 p(X,Y) := g(X)^2 + \sum_{i=1}^r (Y_i^2f_i(X) -1)^2
$$
in the variables $X_1,\ldots,X_n,Y_1,\ldots,Y_r$.  
As in the proof of Lemma~\ref{le:pa-euler} we see that 
$\chi(S)=2^{-r}\chi(\mZ(p))$. 
Let $\rho=2^{2^m}$ be a cone radius of $\mZ(p)$ at infinity 
as in Lemma~\ref{le:bd-cora}. 
Note that $m$ is polynomially bounded in the input size of~$S$
(given by the sparse bit size of the family of polynomials 
describing $S$). 
Consider the semialgebraic set $T\subseteq\R^{n+r+m+1}$ 
defined by 
$$
 p(x,y) =0, z_0=2, z_1 - z_0^2 = 0,\ldots,z_m - z_{m-1}^2 = 0, 
 \|x\|^2 + \|y\|^2 \le z_m^2 .
$$
Clearly, $T$ is homeomorphic to 
$\mZ(p)_\rho = \mZ(p) \cap \{\|x\|^2 + \|y\|^2 \le \rho^2\}$.
Therefore, since $\rho$ is a cone radius, we have 
$\chi(\mZ(p)) = \chi(\mZ(p)_\rho) = \mchi(\mZ(T))$.
By Theorem~\ref{th:euler_sa} we can compute $\mchi(\mZ(T))$ 
in $\FPR^{\CPRi}$. This implies that $\chi(S)$ may be computed 
within the same resources.
\endproof

\begin{remark}\label{re:eebs}
Theorem~\ref{th:euler-bin} easily extends to the case where we 
also allow inequalities $h(x)\ge 0$ in the definition of the basic 
semialgebraic set. For instance, for 
$S=\{x\in\R^n\mid p(x)=0, h(x)\ge 0\}$ consider 
$$
 Z:=\{(x,y)\in\R^{n+1}\mid p(x)=0,h(x)-y^2\} .
$$
The sets $Z_+:= Z\cap\{y\ge 0\}$ and $Z_-:= Z\cap\{y\le 0\}$ are closed 
semialgebraic sets both homeomorphic to $S$ and $Z=Z_+\cup Z_-$. 
The formula
$
 \chi(Z_+\cup Z_-) + \chi(Z_+\cap Z_-) = \chi(Z_+) + \chi(Z_-) 
$
then allows to compute $\chi(S)$ from the Euler characteristic of 
real algebraic varieties.
\end{remark}

\subsection{Connected components and Betti numbers}
\label{se:ccbn}

We are going to study here the following problems: 

\begin{description}
\item{$\CCC$} ({\em Counting connected components})\quad 
Given a semialgebraic set $S$, compute the number of its connected 
components.
\item{$\BETTI{k}$} ({\em $k${\rm th} Betti number of a real algebraic 
set})\quad Given a real multivariate polynomial, compute the $k$th 
Betti number of its real zero set. 

\item{$\MBETTI{k}$} ({\em $k${\rm th} Borel-Moore Betti number of a 
real algebraic set})\quad 
Given a real multivariate polynomial, compute the $k$th Borel-Moore 
Betti number of its real zero set. 
\end{description}

For the problems related to Betti numbers, we restrict the input 
to be a real algebraic set. Since we will only prove lower bounds 
for these problems, this restriction makes our results stronger. 
Note that $\BETTI{0}$ is just the restriction of $\CCC$ to 
real algebraic sets.

We will focus here on the discretized versions of the above problems, 
where the input polynomials have integer coefficients, and study these   
problems in the Turing model. 

The following upper bound was first shown by Canny~\cite{cann:88}. 

\begin{theorem}\label{th:ccc}
The problem $\CCCbit$ is in $\FPSPACE$.
\end{theorem}

From a result by Reif~\cite{reif:79,reif:87} on the $\PSPACE$-hardness 
of a generalized movers problem in robotics, it follows easily that  
the problem $\CCCbit$ is in fact $\FPSPACE$-complete.
We will give an alternative proof of the $\FPSPACE$-hardness of 
this problem following the lines of~\cite{bucu:02}. This will 
also allow us to sharpen the lower bound by showing that 
$\CCCbit$ remains $\FPSPACE$-hard when restricted to compact real algebraic sets. 
Based on this, we will prove the $\FPSPACE$-hardness 
of the problems $\BETTI{k}$ and $\MBETTI{k}$. 

The following lemma follows by inspecting the usual 
$\NPR$-completeness proof of $\FEASR$~\cite{blss:89}. 

\begin{lemma}\label{le:pform} 
For $A\in\PR^0$ there is a polynomial time Turing machine computing 
on input $n\in\N$ a quantifier free first order formula $\Phi_n\in\FRbit$ 
in the free variables 
$x_1,\ldots,x_{p(n)}$ such that the projection 
$$
 \{x\in\R^{p(n)}\mid \Phi_n(x) \mbox{ holds } \} 
 \longrightarrow A\cap\R^n,\ 
 (x_1,\ldots,x_{p(n)})\mapsto (x_1,\ldots,x_n)
$$
is a homeomorphism. The inverse image of an integer point 
$x\in A\cap\Z^n$ is again integer and can be 
computed in polynomial time. 
\end{lemma}

\begin{lemma}\label{le:pequ} 
There is a polynomial time Turing machine computing from  
a quantifier free formula $\Phi\in\FRbit$ in the free variables 
$X_1,\ldots,X_m$ a polynomial $f_\Phi$ in 
$\Z[X_1,\ldots,X_m,Y_1,\ldots,Y_{q(m)}]$ 
such that the projection $\pi\colon\R^{m+q(m)}\to\R^m,(x,y)\mapsto x$ 
induces for all $\epsilon\in\{-1,1\}^{q(m)}$ a homeomorphism
$$
 \mZ(f_\Phi) \cap \{\epsilon_1 y_1\ge 0,\ldots,\epsilon_{q(m)} 
  y_{q(m)}\ge 0\}
  \longrightarrow \{x\in\R^m\mid\Phi(x)\mbox{ holds }\}. 
$$
\end{lemma}

\begin{proof}
As in the $\NPR$-completeness proof of $\FEASR$~\cite{blss:89}
the machine~$M$ performs the following (see also~\cite{cuck:93}).  
For each atomic formula of $\Phi$ containing an inequality  
choose a new variable~$Y$ and replace 
\begin{eqnarray*}
 p(X)\ge 0 &\mbox{ by }& p(X)-Y^2 =0\\ 
 p(X) >  0 &\mbox{ by }& p(X)Y^2-1=0. 
\end{eqnarray*}
In  the resulting formula iteratively eliminate the connectives as 
follows: replace 
$$
\mbox{$\bigvee_{i=1}^s p_i = 0$  by $\prod_{i=1}^s p_i = 0$, and 
      $\bigwedge_{i=1}^t p_i = 0$ by $\sum_{i=1}^t p_i^2=0$.}
$$
We end up with a single polynomial equation $f_\Phi =0$, 
which is easily seen to satisfy the claim of the lemma.
\end{proof}

Consider the following auxiliary problem: 

\begin{description}
\item{$\REACH$} ({\em Reachability})\quad 
Given real polynomials $f,g,h$, decide whether there exist points 
$p\in\mZ_{\R^n}(f,g)$ and $q\in\mZ_{\R^n}(f,h)$ which lie in the 
same connected component of $\mZ_{\R^n}(f)$. 
\end{description}

\begin{proposition}\label{pro:reach}
The problem $\REACHbit$ is $\PSPACE$-hard.
\end{proposition}

\begin{proof}
Assume $L\in\PSPACE$. 
In the proof of~\cite[Proposition~5.9]{bucu:02} 
the configuration graph of a symmetric Turing machine deciding 
membership of $w\in\{0,1\}^n$ to $L$ was embedded in a certain 
way in Euclidean space as a compact one-dimensional 
semi-linear set $S_w$. 
More specifically, a polynomial time computable function mapping
$w\in\{0,1\}^n$ to $(\scC_n,u_n,v_n)$
was constructed, where $\scC_n$ is a constant free additive circuit 
describing membership to $S_w\subseteq\R^{c(n)}$, $c$ is a polynomial, 
and $u_w,v_w\in\{0,1\}^{c(n)}$ such that
$w\in L$ iff $u_w$ and $v_w$ are connected in $S_w$. 
Note that, in particular, the set 
$A:=\{(w,x)\in\{0,1\}^n\times\R^{c(n)}\mid n\in\N, x\in S_w\}$ 
is contained in $\Padd^0$ and hence in $\PR^0$. 

We apply Lemma~\ref{le:pform} to the set $A$. 
Let $\Phi_n\in\FRbit$ be the formula in the free variables
$X_1,\ldots,X_{p(n)}$ corresponding to the input size $n+c(n)$ 
and let $f_n\in\Z[X_1,\ldots,X_{p(n)},Y_1,\ldots,Y_{q(n)}]$ be 
the integer polynomial corresponding to $\Phi_n$ 
according to Lemma~\ref{le:pequ}. We know that 
$f_n$ can be computed from~$n$ in polynomial time.
For $w\in\{0,1\}^n$ let $\mu_w,\nu_w\in\Z^{p(n)}$
be the inverse images of $(w,u_w),(w,v_w)$ under 
the projection homeomorphism 
$$
  T_n:=\{x\in\R^{p(n)} \mid \Phi_n(x)\mbox{ holds }\} 
  \longrightarrow A\cap\R^{n+c(n)},
   (x_1,\ldots,x_{p(n)})\mapsto (x_1,\ldots,x_{n+c(n)}).
$$
Note that $\mu_w$ and $\nu_w$ are connected in $T_n$ iff 
$u_w$ and $v_w$ are connected in $S_w$, which is the case iff $w\in L$. 

According to Lemma~\ref{le:pequ}, for any $\epsilon\in\{-1,1\}^{q(n)}$, 
the projection $(x,y)\mapsto x$ induces a homeomorphism
$$
 \mZ(f_n)\cap \{\epsilon_1 y_1\ge 0,\ldots,
   \epsilon_{q(n)} y_{q(n)}\ge 0\} \longrightarrow T_n.  
$$
This implies that there exist points 
$(\mu_w,\eta),(\nu_w,\eta')\in\mZ(f_n)$ 
that are connected in $\mZ(f_n)$ iff $\mu_w$ and $\nu_w$ 
are connected in $T_n$. 
Define the integer polynomials
$g_w := f_n(\mu_w,Y), h_w := g_n(\nu_w,Y)$.
Then $w\in L$ iff the instance $f_n,g_w,h_w$ 
of the problem $\REACHbit$ has a solution. 
Moreover, $f_n,g_w,h_w$ 
can be computed in polynomial time from $w$. 
\end{proof}

\begin{remark}\label{re:reach1d}
The proof of Proposition~\ref{pro:reach} shows that $\REACHbit$ 
remains $\PSPACE$-hard when restricted to one-dimensional compact 
real algebraic sets.
\end{remark}

\begin{lemma}\label{le:ssusp}
For a compact $Z\subseteq\R^n$ let $\Sigma(Z)\subseteq\R^{n+1}$ 
be the one-point compactification of $Z\times\R$. Then we have 
$b_{\ell+1}(\Sigma(Z))=b_\ell(Z)$ for all $\ell\in\N$. 
(This is also true for $Z=\emptyset $ with the convention that 
$\Sigma(\emptyset)$ is a one point space.)
\end{lemma}

\begin{proof} 
The {\em suspension} $S(Z)$ of a nonempty topological space $Z$ 
is defined as the space obtained from the cylinder 
$Z\times [0,1]$ over $Z$ by identifying the points in each 
of the sets $Z\times \{0\}$ and $Z\times \{1\}$
obtaining the points $v_0$ and $v_1$. 
Essentially, this is a double cone with basis $Z$
and vertices $v_0,v_1$. It is well known that 
the Betti numbers of $S(Z)$ and $Z$ 
are related as follows (cf.~\cite{hatc:02,munk:84}): 
\begin{equation}\label{suspension}
 b_{\ell+1}(S(Z)) = 
 \left\{ \begin{array}{ll} b_\ell(Z)  & \mbox{ if $\ell>0$}\\
                        b_0(Z) -1  & \mbox{ if $\ell=0$.}
         \end{array}\right.
\end{equation}

Assume, without loss of generality, that $Z$ is nonempty. 
Since $Z$~is compact, the one-point compactification $\Sigma(Z)$ of 
$Z\times\R$ is homeomorphic to the suspension of $S(Z)$ where the 
two vertices $v_0$ and $v_1$ of the double cone have been identified.
This space is homotopy equivalent to the space obtained from
the suspension $S(Z)$ by connecting the vertices $v_0$ and $v_1$ 
with a one-dimensional cell.  
This space, in turn, is homotopy equivalent to the space obtained 
from $S(Z)$ by attaching a circle $S^1$ at a point.  
Since this amounts to attach to $Z$ only a cell $e^1$ we conclude 
that 
$$
 b_{\ell+1}(\Sigma(Z)) =  
    \left\{ \begin{array}{ll} 
          b_{\ell+1}(S(Z))  & \mbox{ if $\ell>0$}\\
          b_1(S(Z)) + 1  & \mbox{ if $\ell=0$.}
    \end{array}\right.
$$
Combining this with (\ref{suspension}), the claim
$b_{\ell+1}(\Sigma(Z))=b_\ell(Z)$ follows, 
for any $\ell\in\N$.
%
\end{proof}

The one point compactification of a non-compact 
real algebraic set can
be realized as a real algebraic set by a simple 
construction~\cite[p.~68]{bocr:87}.
For $\xi\in\R^n$ consider the homeomorphism $\iota_\xi$ 
(inversion with respect to the unit sphere with center $\xi$) 
defined by 
$$
 \iota_\xi\colon\R^n-\{\xi\} \longrightarrow\R^n-\{\xi\},\quad 
  x\mapsto \xi + \frac{x-\xi}{\|x-\xi\|^2} .
$$
Let $f$ be a real polynomial of degree $d$ with 
zero set $Z\subseteq\R^n$ and assume that $\xi\not\in Z$. 
Consider the polynomial 
$f^\xi:= \|X-\xi\|^{2d} f(\xi + \|X-\xi\|^{-2}(X-\xi))$
with zero set $Z^\xi\subseteq\R^n$.  
If $Z$ is unbounded then $Z^\xi=\iota_\xi(Z)\cup\{\xi\}$ 
is homeomorphic to the one-point compactification of $Z$. 
Note that if $Z$ is empty,   
then $Z^\xi$ consists just of the point $\xi$. 

\begin{theorem}\label{th:bettis}
For any $k\in\N$ both problems $\BETTIbit{k}$ and $\MBETTIbit{k}$ 
are $\FPSPACE$-hard with respect to Turing reductions.
\end{theorem}

\begin{proof}
Note first that the Borel-Moore and the usual Betti numbers coincide 
for compact sets. 
We denote by $\CBETTIbit{k}$ and $\CREACHbit$ the restrictions of the 
problems $\BETTIbit{k}$ and $\REACHbit$ to compact real algebraic sets. 
We know by Proposition~\ref{pro:reach} and 
Remark~\ref{re:reach1d} that $\CREACHbit$ is $\FPSPACE$-hard. 
To prove the theorem, it is thus sufficient to establish a 
Turing reduction from $\CREACHbit$ to $\CBETTIbit{k}$. 
Our proof is similar to the one of ~\cite[Lemma~5.20]{bucu:02}. 

We first describe a Turing reduction from $\CBETTIbit{0}$ to 
$\CBETTIbit{k}$, for fixed $k>0$.  
Let the compact $Z=\mZ(f)\subseteq\R^n$ be given by 
$f\in\Z[X_1,\ldots,X_n]$.
Set $f_0:=f^2+X_{n+1}^2$, $\xi_0:=(0,\ldots,0,1)\in\R^{n+1}$
and note that $\xi_0\not\in\mZ(f_0)=\mZ(f)\times\{0\}$.  

We recursively compute the sequence of 
polynomials $f_1,\ldots,f_k$ as follows. 
Let $1\le i \le k$ and assume that 
$f_{i-1}\in\R[X_1,\ldots,X_{n+i}]$ 
has already been computed 
such that $\xi_{i-1}:=(0,\ldots,0,1,\ldots,1)\in\R^{n+i}$ 
($n$ zeros, $i$ ones) is not contained in $\mZ(f_{i-1})$. 
Let $\tilde{f}_{i-1}$ denote the polynomial~$f_{i-1}$ interpreted 
as a polynomial in $X_1,\ldots,X_{n+i+1}$, 
where $X_{n+i+1}$ is a new variable and 
$\tilde{\xi}_{i-1}:=(\xi_{i-1},0)\in\R^{n+i+1}$.  
Note that $\mZ(\tilde{f}_{i-1})=\mZ(f_{i-1})\times\R$. 
We define now the polynomial 
$f_{i}:= (\tilde{f}_{i-1})^{\tilde{\xi}_{i-1}}$, which results 
from  $\tilde{f}_{i-1}$ by transformation with the inversion 
$\iota_{\tilde{\xi}_{i-1}}$ w.r.t. the 
unit sphere with center $\tilde{\xi}_{i-1}$ (see the comments before 
Theorem~\ref{th:bettis}). 
Note that $\xi_{i}=\iota_{\tilde{\xi}_{i-1}}(\xi_i)\not\in\mZ(f_i)$ 
since $\|\xi_{i}-\tilde{\xi}_{i-1}\|=1$ and 
$\tilde{\xi}_{i-1}\not\in \mZ(\tilde{f}_{i-1})$.  
Then we have $\mZ(f_{i})=\Sigma(\mZ(f_{i-1}))$ and 
Lemma~\ref{le:ssusp} implies that $b_0(Z)=b_k(\mZ(f_k))$. 
This gives the desired reduction from $\CBETTIbit{0}$ to 
$\CBETTIbit{k}$.

In order to show that $\CREACHbit$ reduces to $\CBETTIbit{0}$ 
we first discuss an auxiliary construction. 
Assume we are given real polynomials $f,g$ such that 
$\mZ(f)\subseteq\R^n$ 
is compact and $\mZ(f,g)$ is nonempty. Consider the one-point 
compactification $Z_{f;g}\subseteq\R^{n+1}$ of the space 
$\mZ(f)\cup (\mZ(f,g)\times\R)$. Topologically, this space is 
obtained from $\mZ(f)$ by attaching a double cone with base $\mZ(f,g)$ 
and identifying the two vertices of this cone. What is 
important is that all the points of $Z(f,g)$ are connected in the 
new space. This is illustrated in Figure~1 below where $\mZ(f)$ is 
the three closed curves, $\mZ(g)$ is the dotted curve and, 
consequently, $\mZ(f,g)$ is the four intersecting points.  
\begin{center}
  \input unr_fig_3.pictex
\end{center}
\begin{center}
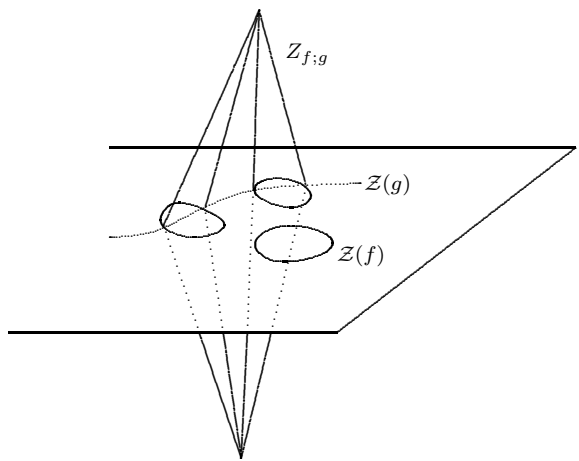
{\small {\bf Figure~1:} An auxiliary construction.}
\end{center}
Using inversions as above, an equation of an algebraic set 
homeomorphic to $Z_{f;g}$ can be easily computed from $f,g$. 
Let $h$ by a further polynomial such that $\mZ(f,h)\ne\emptyset$.
By attaching a double cone with basis $Z(f,h)$ to $Z_{f;g}$,  
we get a real algebraic variety $Z_{f;g,h}$, where all the points of 
$\mZ(f,g)$ and $\mZ(f,h)$, respectively, are connected. 

We describe now the Turing reduction from $\CREACHbit$ to 
$\CBETTIbit{0}$. 
For a given instance $f,g,h\in \Z[X_1,\ldots,X_n]$ of $\CREACHbit$ 
we first check whether $\mZ(f,g)$ or $\mZ(f,h)$ is empty by 
two oracle calls. If this is the case, the corresponding 
reachability problem has no solution. 
Otherwise, we know that both $\mZ(f,g)$ and $\mZ(f,h)$ are 
nonempty. 
We compute now equations for the spaces $Z_{f;g,h}$ and $Z_{f;gh}$  
(note that in the latter, all points of $\mZ(g)\cup\mZ(h)$ have 
been connected). 
The spaces $Z_{f;g,h}$ and $Z_{f;gh}$ have the same number of connected 
components iff  there exist points 
$p\in\mZ(f,g)$ and $q\in\mZ(f,h)$ which lie in the same connected 
component of~$\mZ(f)$. 
Hence we get the desired reduction using two more oracle calls, one 
for $Z_{f;g,h}$ and one for $Z_{f;gh}$. 
\end{proof}

\begin{remark}
The Betti numbers modulo a prime~$p$ are defined similarly 
as the Betti numbers, but replacing the coefficient field $\Q$ 
by the finite field $\F_p$. 
It is easy to check that the proof of Theorem~\ref{th:bettis} 
also gives the $\FPSPACE$-hardness of the 
computation of the $k$th Betti number $\bmod p$,
and similarly for the Borel-Moore Betti numbers. 
\end{remark}

\section{Summary and final remarks}
\label{se:summary}

We have summarized the results of this paper in Figure~2 
which contains three diagrams showing results 
in the Turing model, over $\C$, and over $\R$.  
In this figure, an arrow denotes an inclusion, problems in  
square brackets are Turing-complete for the class at 
their left, problems in curly brackets are 
many-one-complete for that class, and problems in 
angle brackets are hard for that class.  
The problems appearing in the figure are defined in 
the list below.
Recall that if $L$ denotes a problem defined over $\R$ or $\C$, 
we denote its restriction to integer inputs by $L^0$. 

\begin{center}
{\footnotesize 
  \input unr_fig_2.pictex
}
\end{center}
\begin{center}
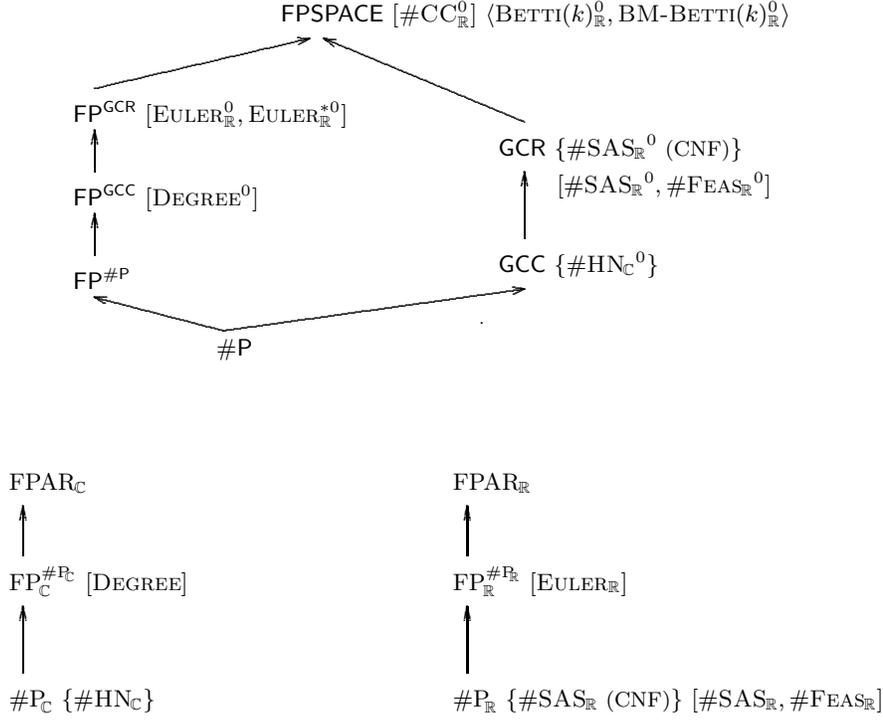
{\small {\bf Figure~2:} Survey of main results.}
\end{center}

\begin{description}\small
\item{$\#\FEASR$} ({\em Real algebraic point counting})\quad 
Given a real multivariate polynomial,
count the number of its real roots, 
returning $\infty$ if this number is not finite. 

\item{$\#\SAS$} ({\em Semialgebraic point counting})\quad 
Given a semialgebraic set~$S$, compute its 
cardinality if $S$ is finite, and return $\infty$ otherwise.

\item{$\#\SAS$ \scriptsize(CNF)} ({\em Semialgebraic point counting})\quad 
Given a semialgebraic set~$S$ in conjunctive normal form, 
compute its cardinality if $S$ is finite, and return $\infty$ 
otherwise.

\item{$\EULER$} ({\em Euler characteristic for basic semialgebraic sets})\quad 
Given a basic semialgebraic set~$S$  
decide whether $S$ is empty and if not, compute $\chi(S)$. 

\item{$\MEULER$} ({\em Modified Euler characteristic})\quad 
Given a semialgebraic set $S$, decide whether it is empty and if not, 
compute its modified Euler characteristic. 

\item{$\CCC$} ({\em Counting connected components})\quad  
Given a semialgebraic set $S$, compute the number of its connected 
components.

\item{$\BETTI{k}$} ({\em $k${\rm th} Betti number of a real algebraic 
set})\quad Given a real multivariate polynomial, compute the $k$th 
Betti number of its real zero set. 

\item{$\MBETTI{k}$} ({\em $k${\rm th} Borel-Moore Betti number of a 
real algebraic set})\quad Given a real multivariate polynomial, 
compute the $k$th Borel-Moore Betti number of its real zero set. 

\item{$\#\HNC$} ({\em Algebraic point counting})\quad 
Given a finite set of complex multivariate polynomials,
count the number of complex common zeros, 
returning $\infty$ if this number is not finite. 

\item{$\DEGREE$} ({\em Geometric degree})\quad 
Given a finite set of complex multivariate polynomials,
compute the geometric degree of its affine zero set. 
\end{description}


Other problems which appeared in 
this paper are listed below. The first three are 
$\NPR$-complete, the other two, $\NPC$-complete. 

\begin{description}\small
\item{$\FEASR$} ({\em Polynomial feasibility})\quad 
Given a real multivariate polynomial, decide whether it has a real root. 

\item{$\SAS$} ({\em Semialgebraic satisfiability})\quad 
Given a semialgebraic set $S$, decide whether it is nonempty.

\item{$\DIMR$} ({\em Semialgebraic dimension})\quad 
Given a semialgebraic set $S$ and $d\in\N$, decide 
whether $\dim S\geq d$.

\item{$\HNC$} ({\em Hilbert's Nullstellensatz})\quad 
Given a finite set of complex multivariate polynomials, 
decide whether these polynomials have a common complex zero. 

\item{$\DIMC$} ({\em Algebraic dimension})\quad 
Given a finite set of complex multivariate polynomials
with affine zero set $Z$ and $d\in\N$, 
decide whether $\dim Z\geq d$.
\end{description}

\begin{remark}\label{re:3repr}
\begin{description}
\item[(i)]
To fix ideas, we assumed in the definition of the above problems  
that the input polynomials are given in sparse representation. 
However, note that choosing the dense encoding leads to  
polynomial time equivalent problems. In order to see this, 
one just has to introduce additional variables that help to 
represent monomials of high degree by ``repeated squaring''. 
The solution set of the new system of polynomial (in)equalities is 
homeomorphic to the original one. A similar remark applies for the 
encoding of polynomials by division free straight-line programs.
%

\item[(ii)] 
Instead of restricting inputs to integer polynomials, 
one could allow also algebraic (or real algebraic) coefficients 
with their standard binary encoding. The results in this 
paper would then hold as well and our proofs would only need 
some extra algorithmics, common in symbolic computation. 
\end{description}
\end{remark}

\section{Open problems}
\label{se:open}

We believe that the developments in this paper 
open up a variety of meaningful new questions. 
To finish this paper we list some of them.

\begin{problem}
Can one decide $\FEASR$ in polynomial time 
with a black box for the Euler characteristic?
\end{problem}

\begin{problem}
It is known that the problem to count the number of 
connected components of a semialgebraic set is 
in $\FPAR$. Is it hard in this class? We know 
that the corresponding result is true 
in the additive setting~\cite{bucu:02}.
\end{problem}

\begin{problem}
What is the complexity to check irreducibility of algebraic 
varieties over~$\C$? And what is the complexity of counting 
the number of irreducible components of algebraic varieties? 
\end{problem}

\begin{problem}
Can Betti numbers of semialgebraic sets be computed in 
$\FPAR$? We know that, in the additive setting, 
the computation of Betti numbers of semi-linear sets 
is $\FPARadd$-complete~\cite{bucu:02}.      
\end{problem}

\begin{problem}
What is the complexity to compute the multiplicity 
$\mathrm{mult}_x(Z)$ of a point $x$ in an algebraic variety~$Z$? 
And how about the computation of intersection 
multiplicities $i(Z,A;x)$? 
\end{problem}


\begin{problem}
What are the Boolean parts $\GCR$ and $\GCC$ of $\CPRi^0$ and $\CPCi^0$,
respectively? 
\end{problem}

\begin{problem}
Toda's theorem~\cite{toda:91} states that $\PH \subseteq \FP^{\CP}$. 
Is there an analogue of this over $\R$ or over $\C$?
\end{problem}


{\footnotesize

}
\end{document}